\documentclass[epj]{svjour}
\usepackage{psfig}

\begin{document}

\title{
From semiconductors to superconductors: a simple model for 
pseudogaps
}
\author{P.~Nozi\`eres and 
F.~Pistolesi\thanks{e-mail: {\tt pistoles@ill.fr}}%
}

\institute{
Institut Laue Langevin BP 156, 38042 Grenoble Cedex 9, France}
\date{Received: / Revised version: }

\abstract{
We consider a two dimensional semiconductor with a local attraction among
the carriers. We study the ground state of this system as
a function of the semiconductor gap. We find a direct transition from a
superconducting to an insulating phase for no doping at a critical value,
the single particle excitations being always gapped.
For finite doping we find a smooth crossover. We calculate the critical
temperature due to both the particle excitations and the
Berezinkii-Kosterlitz-Thouless transition.
}

\maketitle

\section{Introduction}

In the standard BCS theory of superconductors the gap $\Delta $ in the
spectrum of quasiparticles is due to the breakdown of gauge symmetry at the
phase transition, in much the same way as the gap in a spin density wave
state is due to Bragg scattering off the doubled periodicity.\ The gap
disappears with order and consequently $\Delta =1.76\,T_{c}$ is comparable
to the transition temperature $T_{c}$. Largely motivated by high temperature
superconductivity, interest has grown recently in a ``pseudogap'' situation,
in which a somewhat blurred gap $\Delta $ persists in the normal state, such
that $\Delta \gg T_{c}$ \cite{pseudogap}. There are many models for such a
pseudogap, the simplest one being a transition from the BCS limit towards a
``Bose Einstein condensation of preformed pairs'': if the attraction between
electrons is strong enough, bound pairs (let us say singlet) form first,
with a binding energy $\varepsilon _{B},$ then below a critical temperature $
T_{c}$ they Bose condense. It was shown long ago that the evolution between
the two limits is smooth, with however a very different physics in the
strong binding limit \cite{NSR}. Then the gap $\Delta $ is just the binding
energy, a ``molecular'' quantity which has nothing to do with symmetry
breaking; in contrast $T_{c}$ is controlled by center of mass motion of
bound pairs, which are not broken: as a result $\Delta \gg T_{c}$, a simple
minded interpretation of the pseudogap. A similar evolution occurs for the
transition from a paramagnetic metal to an antiferromagnetic insulator in
half filled band: for a strong repulsion the Mott gap $\Delta $ exists
irrespective of the magnetic order.\ The latter is only a secondary feature,
while it was a key actor in the formation of $\Delta $ at 
weak coupling \cite{Mott}.

In order to build an accurate description of realistic systems one needs an
interpolation between these two limits. That means including quantum
fluctuations of the order parameter $\Delta $, (mostly of its phase) which
are systematically ignored in a mean field BCS calculation. That is a
difficult task and the literature on the subject is large \cite{all}. 
We will not attempt to review it: as usual in perturbation 
theories it relies on
approximations that are not really controlled. We will just make a few
simple qualitative comments. Grossly speaking one expects the crossover
between the two limits to occur when the (zero temperature) gap $\Delta $ is
comparable to the Fermi energy $\varepsilon _{F}$: they are the only
energies in the problem. That can be checked explicitly for a two
dimensional system, in which phase fluctuations are enhanced, leading to the
well known Kosterlitz Thouless transition \cite{BKT}. The resulting transition
temperature $T_{c}^{KT}$ is controlled by the phase stiffness of the ground
state: it can be compared to the mean field $T_{c}^{BCS}$ which accounts
only for pair breaking. Such an elementary calculation is sketched in 
section \ref{BKT}: it shows that the crossover $T_{c}^{KT}\approx T_{c}^{BCS}$
occurs when $\Delta \approx \varepsilon _{F}/5$. There a bound pair would
sit in the middle of a continuum, which makes no sense. Sure in the very
strong coupling limit one returns to preformed bound pairs, but there exists
a broad intermediate region near the crossover where phase fluctuations are
already dominant without any bound state.

In the above model the same interaction is responsible for the gap $\Delta $
and for Bose condensation. That may well be the case and the future will
tell whether high temperature superconductors pertain to such a mechanism or
not. However that is not a necessity: one may also imagine that a primary
gap $\Delta _{o}$ is due to some other mechanism that has nothing to do with
superconductivity.\ That may be for instance a lattice distortion of some
sort or a magnetic instability, as suggested by various 
authors \cite{antiferro}: the
starting point is then a semiconductor rather than a 
metal.
If a BCS
attraction between electrons is added to such a system, it may lead to
superconductivity, even if there were no free carriers to start from: free
pairs appear spontaneously if the gain in binding energy is larger than the
cost to produce free electrons and holes across the gap. Such a mechanism is
actually familiar: it was introduced long ago by W. Kohn in his theory of
excitonic insulators \cite{excitonic}. There the Coulomb interaction is
repulsive and the bound pair (the exciton) involves an electron and a hole:
excitons appear spontaneously if their binding energy is larger than the
gap. What we present here is just the reverse situation in which electrons
attract: bound pairs involve two electrons or two holes; at $T=0$ they Bose
condense (bound carriers are then time reversed of each other), hence
superconductivity. It is clear that superconductivity will modify the real
gap $\Delta _{g}$ - hence the name ``primary gap'' for $\Delta _{o}$.

The possibility of superconductivity in a semiconductor is not 
new: it appeared since the beginning of the BCS ``era'',
as emphasized by the review article of M.L. Cohen \cite{cohen}
(for subsequent development, see \cite{hanke}).
However only {\em intraband} interactions in a doped material
were included, the only new feature being the scarcity of carriers.
Here we are concerned with the effect of {\em interband} interactions,
in which carriers in both bands conspire in establishing 
superconducting order. That was also considered before \cite{kohmoto},
but in a somewhat different context of interband pairing, while 
we consider only the usual intraband pairing of time-reversed states.
We show in Appendix \ref{AppB} that the latter is energetically preferred.
(An obvious question is of course whether they can occur together.)

Many features of such a ``superconductivity in a semiconductor''
are general, in the sense that they do not depend on the 
physical mechanism responsible for gap opening. We fell that 
such an interplay of $\Delta_o$ and superconducting order deserves study,
the more as it leads to somewhat unexpected simple conclusions.
The present paper offers a very naive discussion of that problem.
At this stage we have no pretense
whatsoever to explain any realistic material: our only purpose is to
identify and to explain qualitative features of the various limits at hand.
In order to be enlightening and hopefully convincing, such an analysis must
be \textit{simple}: we therefore choose deliberately the most oversimplified
model and formulation, in such a way as to have transparent results.\ We are
perfectly aware of the many loopholes and traps of the model, and we will
try to point those which we feel are important. Starting a discussion of
these many complications would however be meaningless at a stage where even
the feasibility of such a mechanism is an open issue. 

In section \ref{secII} we introduce our model, and we discuss the simplest
situation in which the semiconductor is undoped: were it not for
superconductivity it would be an insulator at $T=0$. Within a mean field BCS
approximation the algebra is elementary and it leads to a direct transition
from insulator to superconductor at zero temperature (when the primary gap $
\Delta _{o}$ reaches a critical $\Delta _{o}^{*}$). At finite $T$ one easily
calculates the order parameter $\Delta$, the actual gap $\Delta _{g}$ and
the critical temperature $T_{c}:$ as could be surmised a pseudogap situation
prevails near $\Delta _{o}^{*}$. Section \ref{secIII} is concerned with
slightly doped systems: then the ground state is always conducting. With an
attraction between carriers it is always superconducting. Still within a BCS
mean field approximation, we explore the various limits.\ The interesting
case is $\Delta _{o}>\Delta _{o}^{*}$, where everything is due to doping;
the behaviour of gaps and critical temperature in that limit is somewhat
unexpected. Up to that level the dimension of space and the nature of the
gap in the Brillouin zone do not matter much: they become crucial in section
\ref{BKT} which is concerned with a much more delicate issue, namely phase
stiffness and Kosterlitz Thouless fluctuations. The analysis is here more
speculative: as far as we can see the transition is always monitored by
phase fluctuations in realistic models for the gap, in contrast to ordinary
BCS metals.\ The reason is a very small phase stiffness, which would vanish
for a filled band. In a brief conclusion we emphasize the many questions
that remain open.

\section{The BCS model for an undoped semiconductor}

\label{secII}

\begin{figure}[tbp]
\centerline{\psfig{file=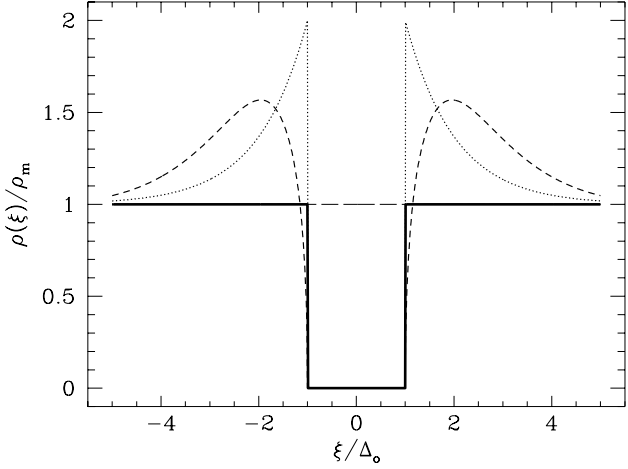,width=9cm,angle=-90}}
\caption{Possible modification of the density of states near the band edge. 
The thick line correspond to the approximation used in the text. 
The dotted line is instead an example of a more realistic 2d case 
where the removed stated remain near the edge. The dashed line is 
a typical 3d case where the density of states vanishes at the edge
with a square root dependence on the energy.}
\label{dos}
\end{figure}

We start from a normal Fermi liquid that has a gap $2\Delta _{o}$, and we
choose the origin of energy at midgap. The gap may be direct or indirect in
momentum space: at that stage it does not matter. Let $\rho(\xi)$ 
be the density of states (per spin): in a first approximation we
take $\rho(\xi)$ constant: everything is symmetric around 
$\xi =0$. Since the total number of states must remain the same, opening
the gap $\Delta _{o}$ means that the conduction band shifts from 
$\left[0,\omega _{m}\right]$ to 
$\left[ \Delta _{o},\Delta _{o}+\omega _{m}\right]$ 
while the valence band goes to 
$[ -\Delta _{o}-\omega _{m},-\Delta_{o}]$. 
The conduction and valence bands are just shifted away from
each other. $\omega _{m}$ is a bandwidth which we assume $\gg \Delta _{o}$
(the gap is small). The choice of a constant $\rho $ is in accordance with
our choice of simplicity, but it is not realistic for two reasons 
(\emph{i\/}) 
at band edges $\rho $ goes as $\xi^{1/2}$ in three dimensions, 
(\emph{ii\/}) 
in usual situations states that are repelled off the gap remain
in its vicinity, in an energy range which has the same order of magnitude $
\Delta _{o}$ (see Fig.~\ref{dos}); here these states are repelled at
infinity.
We will show in Appendix \ref{AppA} that more realistic choices for 
$\rho(\xi)$ do not change the results qualitatively: they are not
worth the price of complications.

\begin{figure}[tbp]
\centerline{\psfig{file=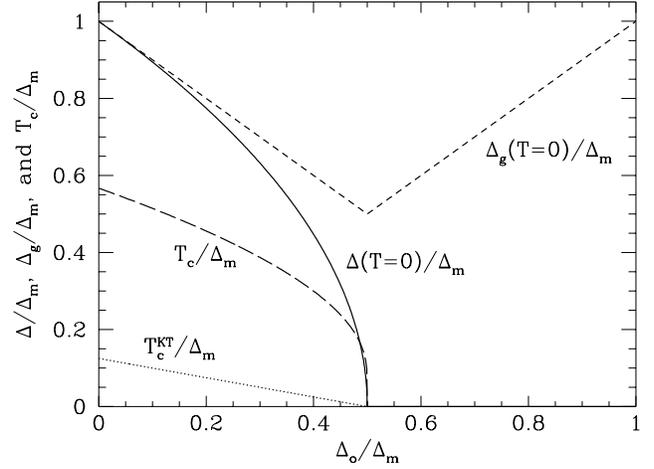,width=9cm,angle=-90}}
\caption{Undoped case, we report as a function of $\Delta_o/\Delta_m$ the 
following quantities: the order parameter $\Delta /\Delta _{m}$ 
(continuous line), the excitation gap $\Delta _{g}/\Delta_{m}$ (dashed line), 
the critical temperature $T_c/\Delta_m$ 
due to pair breaking (long dashed line), 
and the Kosterlitz-Thouless critical temperature $T_c^{KT}/\Delta_m$, 
[see section \ref{BKT}] (dotted line).
}
\label{undoped}
\end{figure}

We then add a local attraction $U$ between fermions and we look for a BCS
superconducting state in which electrons (or holes) are paired in singlet
states with zero total momentum, $\left( {\bf k} \uparrow \right) $ 
with $\left(-{\bf k}\downarrow \right)$ (one the time reversed of the other). 
If the bottom of the band is made up of several
valleys at band edge, that means pairing of electrons between symmetric
valleys. We treat the problem within a mean field approximation: the
Hamiltonian contains anomalous terms
\[
\Delta \,c_{{\bf k}\uparrow }\,c_{-{\bf k}\downarrow }
\]
where $\Delta $ is the usual order parameter, not to be confused with gaps,
which must be determined self-consistently. Because of electron-hole
symmetry the chemical potential $\mu $ of our undoped system remains at
midgap and the reduced kinetic energy $\xi _{k}=\left( \varepsilon _{k}-\mu
\right) $ starts at $\pm \Delta _{o}$ instead of 0. The algebra is identical
to that of BCS except for that shift in the energy integration. In the
absence of a gap, $\Delta _{o}=0$, the order parameter $\Delta $ would take
the regular metallic value $\Delta _{m}$ given by the familiar BCS self
consistency condition
\begin{equation}
1={\frac{\rho U}{2}}\int_{-\omega _{m}}^{+\omega _{m}}\frac{d\xi }{\sqrt{\xi
^{2}+\Delta _{m}^{2}}}=\rho U\,\,\log \frac{2\omega _{m}}{\Delta _{m}}\,.
\end{equation}
It is convenient to characterize the attraction by $\Delta _{m}$ instead of $
U$. Note that for a regular metal the superconducting gap is just $
\Delta _{g}=\Delta $. In the presence of a semiconductor gap $\Delta _{o}$
the self consistency equation for the \textit{new} order parameter $\Delta $
becomes
\begin{equation}
1=\rho U\,\log \left[ \frac{2\omega _{m}}{\Delta _{o}+\sqrt{\Delta
_{o}^{2}+\Delta ^{2}}}\right] \,.
\end{equation}
In order for the two equations to have a solution with the same $U$, 
we must have $\Delta _{m}=\Delta _{o}+\sqrt{\Delta _{o}^{2}+\Delta ^{2}}$,
{\em i.e.}
\begin{equation}
\Delta =\sqrt{\Delta _{m}\left( \Delta _{m}-2\Delta _{o}\right) }\,.
\label{eq3}
\end{equation}
Superconductivity persists with a reduced order parameter until the gap $
2\Delta _{o}$ reaches the metallic BCS gap $\Delta _{m}$: then the gain in
interaction energy due to superconductivity can no longer overcome the cost
of kinetic energy in producing free carriers across the gap. Note the
complete analogy with excitonic insulators in which condensation produces
electron-hole pairs in the presence of a repulsive $U$ \cite{excitonic}.
Such a model provides an elementary example of a direct \textit{\
superconductor-insulator transition}, which here is sharp at $T=0$. The
quasiparticle gap $\Delta _{g}$ is just $\Delta _{o}$ in the insulating
state ($\Delta =0$). In the superconducting state it is $\sqrt{\Delta
_{o}^{2}+\Delta ^{2}}$, \emph{i.e.} $\Delta _{g}=\Delta _{m}-\Delta _{o}$.
The situation is illustrated on Fig.~\ref{undoped}: note that the gap is
finite when superconductivity disappears.

The calculation is easily extended at finite temperatures. The order
parameter $\Delta $ becomes a function of $T$. We need only introduce the
fermionic occupation factors in the gap equation:
\begin{equation}
\int_{P}d\xi {\frac{\tanh \beta \sqrt{\xi ^{2}+\Delta ^{2}}/2}{\sqrt{\xi
^{2}+\Delta ^{2}}}}=\int_{-\omega _{m}}^{\omega _{m}}{\frac{d\xi }{\sqrt{\xi
^{2}+\Delta _{m}^{2}}}}\,.  \label{eq13}
\end{equation}
(we retain the zero temperature $\Delta _{m}$ as a measure of the attraction
$U$). The integration path $P$ takes into account the absence of states in
the middle of the band: $P=[-(\Delta _{o}+\omega _{m}),-\Delta _{o}]$ and $
[+\Delta _{o},+(\omega _{m}+\Delta _{o})]$. For $\Delta _{o}=0$,  Eq.~(\ref
{eq13}) is the standard BCS gap equation, yielding the 
usual $\Delta (T)$ and $T_{c}$. 
At the other end the critical temperature goes to zero at the
superconductor-insulator transition $\Delta _{o}^{*}=\Delta _{m}/2$ (there $
\Delta =0$ at zero temperature: there is no need of a $T_{c}$ to make it
vanish). The behaviour of $T_{c}(\Delta _{o})$ is shown on Fig.\ 2.

The approach to the transition is interesting: we subtract the $\Delta
_{o}<\Delta _{o}^{*}$ Eq.~(\ref{eq13}) from the same equation at the
transition, $\Delta _{o}=\Delta _{o}^{*}$:
\begin{equation}
\int_{\Delta _{o}^{*}}^{+\infty }{\frac{1-\tanh \beta _{c}\xi /2}{\xi }}d\xi
-\int_{\Delta _{o}}^{\Delta _{o}^{*}}{\frac{\tanh \beta _{c}\xi /2}{\xi }}
d\xi =0\,.
\end{equation}
When $\Delta _{o}\rightarrow \Delta _{o}^{*}$, $T_{c}$ vanishes and we can
approximate $\tanh x$ with $1-2e^{-2x}$. The first integral reduces then to
the Exponential integral function and for large values of $\beta _{c}\Delta
_{o}^{*}$ we obtain:
\begin{equation}
2{\frac{T_{c}}{\Delta _{o}^{*}}}e^{\Delta _{o}^{*}/T_{c}}={\frac{\Delta
_{o}^{*}-\Delta _{o}}{\Delta _{o}^{*}}}\equiv \tau \,.  \label{eqcrit}
\end{equation}
Defining the function $w(x)$ as the solution of the transcendental equation $
x=w\,e^{w}$, we can write $T_{c}$ in terms of $w(x)$
\begin{equation}
T_{c}={\frac{\Delta _{o}^{*}}{w(2/\tau )}}\,.
\end{equation}
The function $w(x)$ is a generalization of the logarithm and for $x\gg 1$
has the following expansion:
\begin{equation}
w(x)=\ln x-\ln [\ln x-\ln [\ln x-\ln \dots ]]\,.  \label{dev}
\end{equation}
The convergence of that expansion is very slow [shown in (\ref{dev})], but
the first term is enough to provide a rough result: the critical temperature
vanishes logarithmically in $\tau $ [$T_{c}\sim \Delta _{o}^{*}/\ln (2/\tau
) $].

\begin{figure}[tbp]
\centerline{\psfig{file=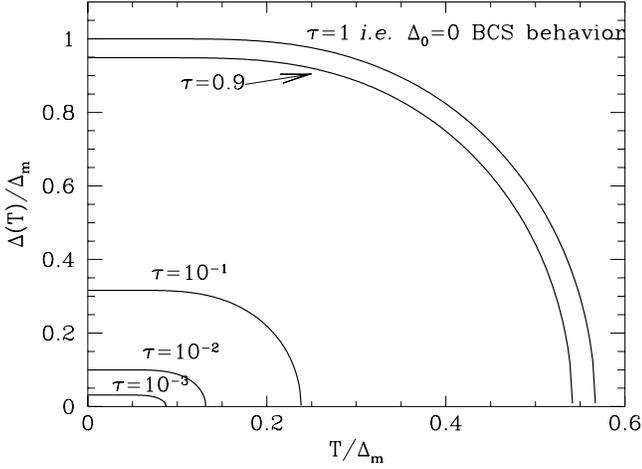,width=9cm,angle=-90}}
\caption{$\Delta (T)$ for $\tau=1$, 0.9, $10^{-1}$, $10^{-2}$ and 
$10^{-3}$. The evolution from the familiar BCS shape for large $\tau$ to 
a new one for $\tau$ small is evident.}
\label{deltaT}
\end{figure}

Altogether the presence of a semiconductor gap $\Delta _{o}$ acts to
decrease $T_{c}$, a result which is reasonable. But such a decrease should
be compared to that of the zero temperature order parameter $\Delta $, which
behaves as $\sqrt{\tau }$: we see that the ratio $\Delta (T=0)/T_{c}$
becomes extremely small when $\Delta _{o}\rightarrow \Delta _{o}^{*}$, in
contrast to the usual BCS result. The physical origin of that behaviour is
clear: due to the finite gap $\Delta _{g}$ the quasi particle excitations
are much less efficient in destroying the superconducting order than they
would be in the BCS case, and consequently $T_{c}$ increases. Such a
behaviour is exemplified in Fig.~\ref{deltaT} in which we plot the order
parameter $\Delta (T)$ [obtained from solving (\ref{eq13})] for various
values of $\Delta _{o}/\Delta _{m}$: the elongated low plateau near the
transition is striking.

All that discussion is based on a BCS like mean field approximation, in
which condensed pair breaking is the only allowed thermal excitation,
ultimately responsible for the destruction of superconductivity at $T_{c}$.
Such an approximation is by no means obvious and it should be kept in mind.
In practice there exist other excitations that might be more efficient in
destroying superconductivity. In particular, phase fluctuations are gapless
excitations and at least for $\Delta _{o}$ near $\Delta _{o}^{*}$ they are
expected to reduce drastically the critical temperature. We will discuss
this point more extensively in section \ref{BKT} for the two-dimensional
system. There we find that phase fluctuations seem to be always responsible
for the transition to the normal phase, even for $\Delta _{o}=0,$ a somewhat
unexpected conclusion.

\section{Doped system}

\label{secIII}

We now extend our problem, allowing for a finite doping.\ We do it because
it leads to new interesting physics: the sharp transition at $\Delta
_{o}^{*} $ becomes a smooth crossover, and superconductivity persists at
zero temperature for any value of $\Delta _{o}$. Beyond that crossover, the
superconducting behaviour is very unconventional over a broad range of
parameters, in contrast to the undoped case in which queer behaviour
occurred only at a specific transition point. In that respect the doped
system is much more generic.

\begin{figure}[tbp]
\psfig{file=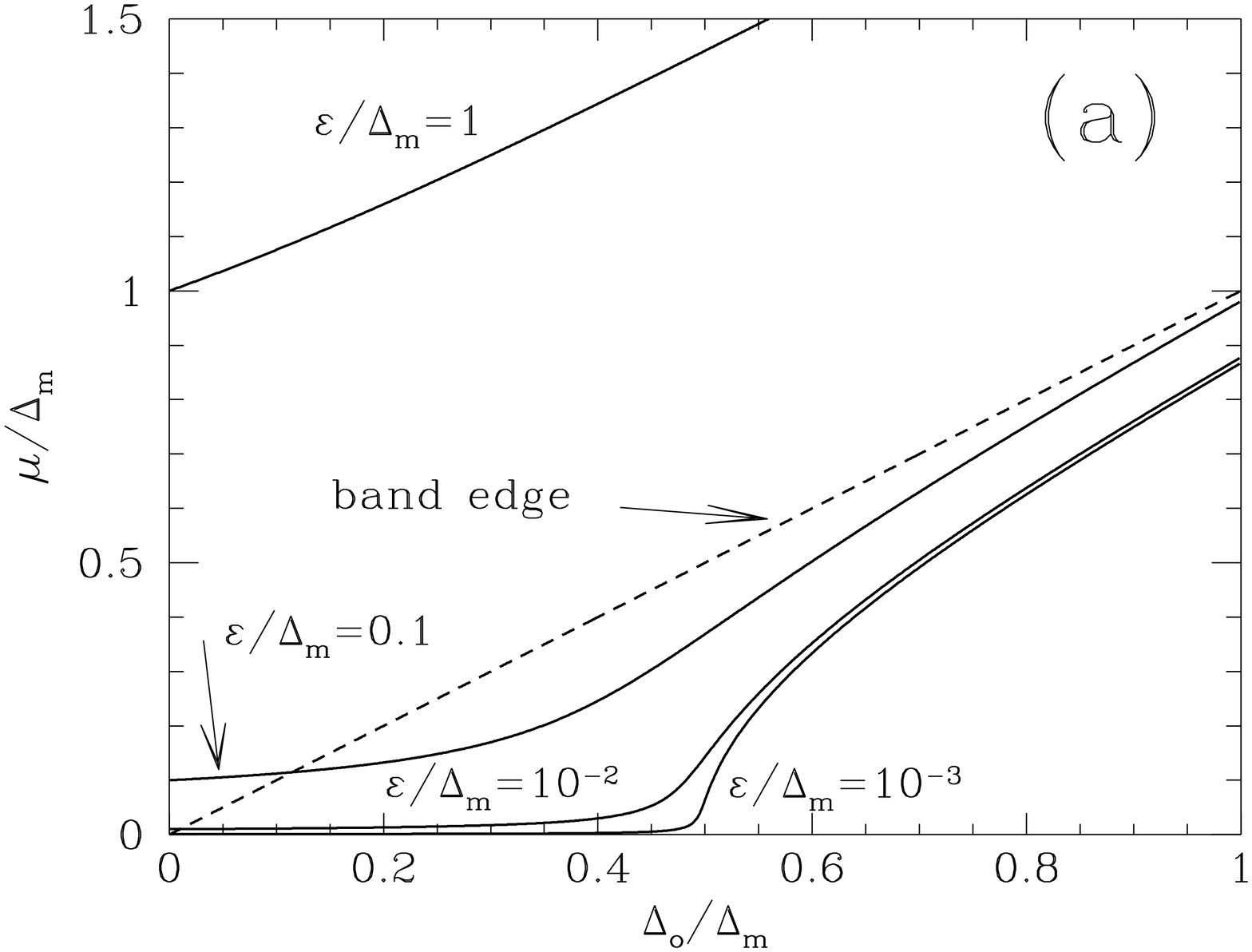,width=9cm,angle=-90}
\psfig{file=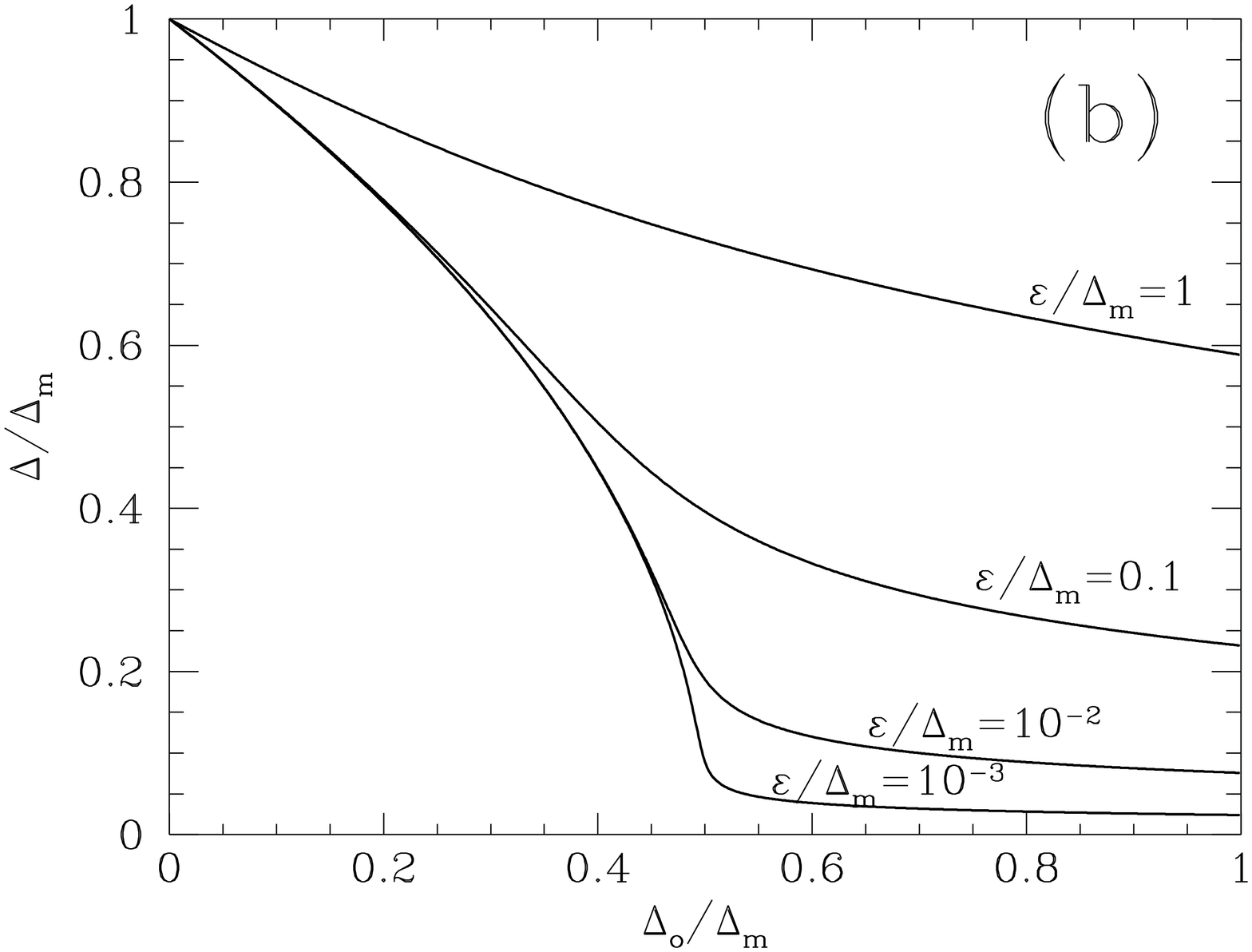,width=9cm,angle=-90}
\caption{ (a) Chemical potential ($\mu $) for different value of $
\varepsilon /\Delta _{m}$ as a function of $\Delta _{o}/\Delta _{m}$. 
The dashed line is
the bottom of the upper band ($\Delta _{o}/\Delta _{m}$).
(b) The same for the order parameter $\Delta /\Delta _{m}$. 
 }
\label{fig2}
\end{figure}

Starting from the non interacting, metallic system at $T=0$ we suppose that
the upper band is populated up to an energy $\mu _{m}=\varepsilon $,
referred to midgap. When the gap $\Delta _{o}$ is restored, the
''semiconductor'' is doped by an amount $\delta N=2\rho \varepsilon $
(referred to the filled valence band). In the metallic state 
$\Delta _{o}=0$, chemical potential and 
superconductivity are unaffected by each other (the
BCS picture is symmetric with respect to Fermi level): $\mu _{m}$ remains
equal to $\varepsilon $. In the opposite limit of an insulator with no
superconductivity, $\mu =\Delta _{o}+\varepsilon $ is pushed into the
conduction band, a familiar feature of degenerate semiconductors. In between
we expect no sharp superconductor-insulator transition: a weak
superconducting order due to the few free carriers in the conduction band
should persist at any $\Delta _{o}$. Keeping $\Delta _{m}$ and $\varepsilon $
fixed, we thus expect the zero temperature chemical potential $\mu $ and
order parameter $\Delta $ to depend on $\Delta _{o}$ as showed respectively
on Fig.~\ref{fig2}~(a) and (b). The transition around $\Delta _{o}=\Delta
_{m}/2$ is a crossover, but with a large change of $\mu $.

Let us assume an arbitrary temperature $T$. In order to fix the two
parameters $\mu (T)$ and $\Delta (T)$ we need two conditions. One is the
self consistency equation for the order parameter, which is the same as (\ref
{eq13}) but for the introduction of the chemical potential in the expression
for the quasiparticle energy:
\begin{equation}
\int_{P}\!\!{\frac{\tanh \beta E(\xi )/2}{E(\xi )}}d\xi  =\int_{-\omega
_{m}}^{+\omega _{m}}{\frac{d\xi }{\sqrt{\xi ^{2}+\Delta _{m}^{2}}}}\,,
\label{eqA}
\end{equation}
where $E(\xi )=\sqrt{(\xi -\mu )^{2}+\Delta ^{2}}$. The other one is the
conservation of particle number, stating that $\delta N$ is unchanged:
\begin{equation}
2\rho \int_{P}d\xi \left[ v^{2}(\xi )+f_{F}[E(\xi )]{\frac{\xi -\mu }{E(\xi )%
}}\right] -2\rho \int_{-\omega _{m}}^{0}d\xi =\delta N \,,
\label{eqB}
\end{equation}
where $f_{F}(E)=1/(e^{\beta E}+1)$ is the Fermi factor for inverse
temperature $\beta $ and
\begin{equation}
v^{2}(\xi )=1-u^{2}(\xi )={\frac{1}{2}}\left[ 1-{\frac{\xi -\mu }{E(\xi )}}
\right]
\end{equation}
is the familiar BCS occupation factor. The cut-off $\omega _{m}$ can be sent
to infinity in Eqs. (\ref{eqA})-(\ref{eqB}): we are left with a system of
two coupled equations.

\subsection{Ground state properties}

We begin by considering the zero temperature limit. In this case
integrations are elementary and we obtain:
\begin{equation}
\left\{
\begin{array}{rcl}
\Delta _{m}^{2} & = & \left[ \Delta _{o}-\mu +\sqrt{\left( \Delta _{o}-\mu
\right) ^{2}+\Delta ^{2}}\right] \cdot 
	\left[ \Delta _{o} +\mu 
\vphantom{\sqrt{\left(\Delta _{o}+\mu \right) ^{2}+\Delta ^{2}}}
	   \right.
	\\
	&& \qquad\left.  +\sqrt{\left(
	\Delta _{o}+\mu \right) ^{2}+\Delta ^{2}}\right] \\
&  &  \\
2\varepsilon & = & 2\mu -\sqrt{(\Delta _{o}+\mu )^{2}+\Delta ^{2}}+\sqrt{
(\Delta _{o}-\mu )^{2}+\Delta ^{2}}
\end{array}
\right. \,.  \label{system0}
\end{equation}
System (\ref{system0}) fixes $\mu $ and $\Delta $ as a function of the gap $
\Delta _{o}$ and doping $\delta N=2\rho \varepsilon $. Once they are known,
the overall gap for elementary excitations is either $\Delta $ (if $\mu
>\Delta _{o})$ or $\sqrt{\left( \Delta _{o}-\mu \right) ^{2}+\Delta ^{2}}$
(if $\mu <\Delta _{o}$). Numerical solution of these equations is easy and
it is shown in Fig. \ref{fig2}. Simple limits can be assessed analytically
as long as $\varepsilon \ll \Delta _{m}$, shedding some light on the
physical meaning of these results.

Consider first the ``conducting'' side. Here free carriers extend deep into
the valence and conduction bands: the chemical potential for small doping
will remain close to midgap. We can therefore expand the second equation of 
(\ref{system0}) into
\begin{equation}
\varepsilon =\mu \left[ 1-\frac{\Delta _{o}}{\sqrt{\Delta _{o}^{2}+\Delta
^{2}}}\right] \,.  \label{eq8}
\end{equation}
Doping is a small perturbation and we can use the undoped result for $\Delta
$: we thus find
\begin{equation}
\mu =\frac{\Delta _{m}-\Delta _{o}}{\Delta _{m}-2\Delta _{o}}\,\varepsilon
+O(\varepsilon ^{2})\,.  \label{eq9}
\end{equation}
The approximation breaks down when $\varepsilon /\Delta _{o}^{*}\sim \tau $.
The crossover is very close to the super\-con\-duc\-tor-insulator sharp
transition found when $\delta N=0$. Substituting that first order result for
$\mu $ into the first of (\ref{system0}) we find the correction to 
$\Delta$:
\begin{equation}
\Delta =2\Delta _{o}^{*}\sqrt{\tau }+\left( {\frac{\varepsilon }{\Delta
_{o}^{*}}}\right) ^{2}{\frac{1+\tau }{8\,\tau ^{5/2}}}\Delta
_{o}^{*}+O(\varepsilon ^{4})\,.
\end{equation}
As long as $\Delta _{o}<\Delta _{o}^{*}$ doping is a minor perturbation.

In the opposite limit, we first note that an insulating state at $T=0$ is
impossible. It would mean a $\Delta =0$ solution to the gap equation, 
{\em i.e.} $\Delta _{m}^{2}=4(\Delta _{o}^{2}-\mu ^{2})$, which in turn
implies $\mu <\Delta _{o}$. But with no $\Delta $ that means $\delta N=0$.
Thus superconductivity must either extend all the way, or display a first
order transition: a sharp second order transition to the insulating state is
precluded. We now show that the present model has a smooth crossover.

Deep in the insulating limit, $\Delta _{o}>\Delta _{o}^{*},$ the role of
equations is interchanged. The order parameter $\Delta $ will turn out to be
small: consequently the gap equation which used to fix $\Delta $ now fixes $
\mu $
\begin{equation}
\Delta _{m}^{2}=4(\Delta _{o}^{2}-\mu ^{2})\quad \Rightarrow \quad \mu =%
\sqrt{\Delta _{o}^{2}-{\Delta _{o}^{*}}^{2}}\,.  \label{eq10}
\end{equation}
Such an expansion is valid as long as $\left( \Delta _{o}-\mu \right) \gg
\Delta $, which implies
\begin{equation}
\frac{{\Delta _{o}^{*}}^{2}}{2\Delta _{o}}\gg \Delta
\label{condition}
\end{equation}
$\mu $ is still inside the band gap $\Delta _{o}$.\ Granted $\mu ,$ the
other equation for $\delta N$ fixes the order parameter $\Delta $
\begin{equation}
\varepsilon =\frac{\delta N}{2\rho }=\frac{2\Delta ^{2}}{{\Delta _{m}}^{2}}%
\sqrt{\Delta _{o}^{2}-{\Delta _{o}^{*}}^{2}}\quad \Rightarrow \quad \Delta =%
\frac{\Delta _{m}\sqrt{\varepsilon /2}}{\left[ \Delta _{o}^{2}-{\Delta
_{o}^{*}}^{2}\right] ^{1/4}}\,.  \label{eq12}
\end{equation}
As surmised $\Delta $ is small for small doping $\varepsilon $, which
justifies our calculation a posteriori. The condition (\ref{condition}) 
now reads
\begin{equation}
\varepsilon \ll \frac{{\Delta _{o}^{*}}^{2}}{8\Delta _{o}} \,.
\label{cond2}
\end{equation}
Doping should remain small (or the gap $\Delta _{o}$ not exceedingly large).

When the gap exceeds that range the expansion in powers of $\Delta $ is no
longer warranted. But as long as doping $\varepsilon $ is small compared to $%
\Delta _{o}^{*}$ we expect the chemical potential to remain close to $\Delta
_{o},$ either below (inside the gap) or above (inside the continuum). We may
then assume safely that $\left( \Delta _{o}+\mu \right) \approx 2\Delta
_{o}\gg \Delta .$ The two basic equations (\ref{system0}) simplify into
\begin{equation}
\left\{
\begin{array}{rcl}
\Delta _{o}^{*2} 
	& = & 
\left[ \Delta _{o}-\mu +\sqrt{\left( \Delta _{o}-\mu\right)^{2}
	+\Delta ^{2}}\right] \cdot \Delta _{o} 
\\
	2\varepsilon  & = & 
\mu -\Delta _{o}+\sqrt{(\Delta _{o}-\mu )^{2}+\Delta ^{2}} 
\end{array}
\right.
\label{system2}
\end{equation}
Multiplying one by the other we obtain the very simple result
\[
\Delta ^{2}=\frac{2\varepsilon \Delta _{o}^{*2}}{\Delta _{o}}
\]
which joins smoothly with our former expansion (\ref{eq12}), 
and which remains valid throughout the crossover 
$\left| \Delta_{o}-\mu \right| \sim \Delta$. 
Similarly we may eliminate the square roots
from these two equations, which yields the chemical potential $\mu $%
\begin{equation}
\mu =\Delta _{o}+\varepsilon -\frac{\Delta _{o}^{*2}}{2\Delta _{o}}
\,.
\end{equation}
For very large gaps $\Delta_o\gg{\Delta_o^*}^2/\varepsilon$,
 the last term is negligible and we
recover the usual result for a doped normal semiconductor, $\mu =\Delta
_{o}+\varepsilon $.

Let us explore these two regimes in some more detail. When the chemical
potential lies inside the gap $\Delta _{o}$ the real gap $\Delta _{g}$ is
just $\left( \Delta _{o}-\mu \right) \gg \Delta $, of order $\Delta
_{o}^{*2}/2\Delta _{o}$ as $\mu $ approaches the band edge. The relevant
ratio is
\begin{equation}
\frac{\Delta }{\Delta _{g}}=\sqrt{\frac{8\varepsilon \Delta _{o}}{\Delta
_{o}^{*2}}}\ll 1 \,.
\end{equation}
Note the strong similarity between such a regime and the Bose condensation
of preformed pairs \cite{NSR}. In the latter case the chemical potential $\mu
$ is close to the energy of bound pairs, $\varepsilon _{b}/2$ per fermion
and therefore it lies inside the gap. Here the reduced kinetic energy ($\xi
-\mu $) likewise never vanishes: $\left( \Delta _{o}-\mu \right) $ plays the
role of $\varepsilon _{b}/2$. The resulting behaviour is similar in both
cases, with a completely different physics. Here the hierarchy of energy
scales is
\[
\varepsilon \ll \Delta \ll \Delta _{g}
\]
(indeed one has $\Delta ^{2}=4\varepsilon \Delta _{g}$). In the opposite
limit the chemical potential enters the conduction band: then the real gap $
\Delta _{g}$ is just the order parameter $\Delta .$ The relevant ratio is
that of $\Delta $ to the Fermi energy $\varepsilon $ measured from the bottom
of the band
\[
\frac{\Delta }{\varepsilon }=\sqrt{\frac{2\Delta _{o}^{*2}}{\varepsilon
\Delta _{o}}}
\]
Except for a numerical factor 4 we see that $\Delta \ll \varepsilon $: we
recover an ordinary BCS regime in which superconducting pairing is a small
perturbation.

What is unusual is the behaviour of the order parameter $\Delta \approx
\sqrt{\varepsilon }$ in that ``insulating regime'', for small doping $
\varepsilon $ and moderate semiconductor gap $\Delta _{o}$. Such an unusual
dependence $\Delta \approx \sqrt{\varepsilon }$ is due to the interplay of
doping and superconductivity: 
({\em i\/}) extrinsic carriers are needed in order to
start superconductivity, 
but ({\em ii\/}) the latter does add further carriers,
thereby modifying $\Delta$. That result for the dependence on doping away
from ${\Delta _{o}^{*}}$ should be compared with the result $\Delta =2\Delta
_{o}^{*}\sqrt{\tau }$ for the undoped system close to ${\Delta _{o}^{*}}$.
The two variables $\tau $ and $\varepsilon $ act in the same way in driving
the system away from the $T=0$ critical point (we will find this behaviour
again for the critical temperature).

Analytic calculations are possible only in limiting cases.\ In the crossover
region one must resort to numerical solutions, which are quite
straightforward. The result is shown in Fig.~\ref{fig2} where system (\ref
{system0}) is solved numerically for the values of $\varepsilon $ indicated.
Both $\Delta $ and $\mu $ clearly display the transition from a ``metallic''
to a ``semi insulating'' regime. In the latter, $\mu $ crosses the band edge
when $\varepsilon \sim \Delta _{o}^{*2}/2\Delta_o $, separating Bose Einstein
like and BCS like behaviours. Note that the chemical potential crosses the
band edge $\Delta _{o}$ twice when $\varepsilon $ is smaller than a critical
value (see also the discussion after Fig.~\ref{phase}).

\subsection{Finite temperature behaviour}

\begin{figure}[tbp]
\centerline{\psfig{file=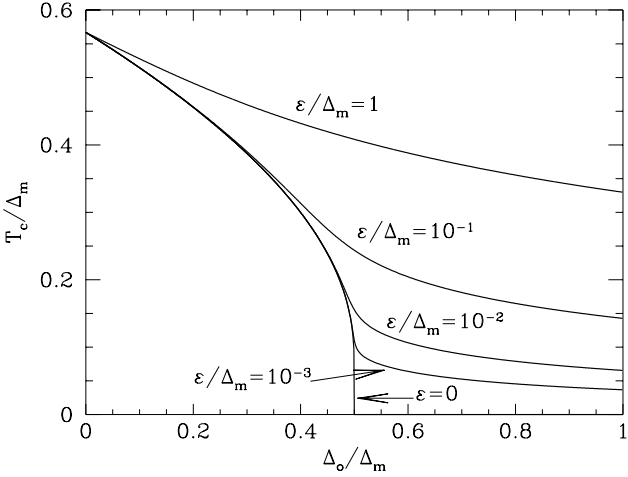,width=9cm,angle=-90}}
\caption{Critical temperature $T_{c}$ as a function of $\Delta _{o}/\Delta
_{m}$ for $\varepsilon /\Delta _{m}=$ $0$, $10^{-3}$, $10^{-2}$, $10^{-1}$, and
$1$.}
\label{fig3}
\end{figure}

Having a complete view of the zero temperature doped system, we consider now
the effect of a finite temperature. In particular we want to investigate the
critical temperature of the system. Setting $\Delta =0$ into 
Eqs. (\ref{eqA})-(\ref{eqB}) we obtain:
\begin{eqnarray}
\int_{\Delta _{o}}^{\infty }\!\!d\xi 
\left[ 
\vphantom{\frac{2}{\sqrt{\xi ^{2}+\Delta _{m}^{2}}}}
{\frac{\tanh \beta _{c}(\xi -\mu _{c})/2}{
\xi -\mu _{c}}}+{\frac{\tanh \beta _{c}(\xi +\mu _{c})/2}{\xi +\mu _{c}}}
\right.  && \nonumber \\
\left. -{%
\frac{2}{\sqrt{\xi ^{2}+\Delta _{m}^{2}}}}\right] 
	=2\int_{0}^{\Delta
_{o}}\!\!\!\!{\frac{d\xi }{\sqrt{\xi ^{2}+\Delta _{m}^{2}}}}
& &  \label{Tceq1}
\end{eqnarray}
and
\begin{equation}
\varepsilon =2\mu _{c}+{\frac{1}{\beta _{c}}}\log \left[ {\frac{1+e^{\beta
_{c}(\Delta _{o}-\mu _{c})}}{1+e^{\beta _{c}(\Delta _{o}+\mu _{c})}}}\right]
\,.  \label{Tceq2}
\end{equation}
These equations can be easily solved numerically and the results are shown
in Fig.~\ref{fig3} for different values of $\varepsilon $ including the
undoped case.
As expected from the zero temperature analysis, the critical temperature
also exhibits a smooth crossover from a large to a small value, and it never
vanishes for finite doping. The region $\Delta _{o}<\Delta _{o}^{*}$ is not
much different from a standard BCS superconductor: a comparison of Fig.~\ref
{fig3} with Fig.~\ref{fig2}~(a) clearly shows that $T_{c}$ is proportional
to $\Delta (T=0)$ (with the same BCS factor for $\Delta \rightarrow 0$).

A new and interesting behaviour is instead found when $\Delta _{o}>\Delta
_{o}^{*}$. In this region the critical temperature appears to be much larger
than the order parameter $\Delta (T=0)$ when $\varepsilon /\Delta _{m}\ll 1$
. This point can easily be verified analytically. In this limit we can
substitute the ground state chemical potential $\mu _{o}$ in Eq.~(\ref{Tceq2}
). Since the critical temperature vanishes for $\varepsilon =0$ while $
\Delta _{o}-\mu _{o}>0$ the exponentials in Eq.~(\ref{Tceq2}) are large and
we can expand the expressions. At leading order in $\exp \{(\mu _{o}-\Delta
_{o})/2T_{c}\}$ we obtain the following equation for $T_{c}$:
\begin{equation}
T_{c}=\varepsilon \,e^{\frac{\mu _{o}-\Delta _{o}}{T_{c}}}\,.
\end{equation}
Again the solution can be written in terms of $w(x)$:
\begin{equation}
T_{c}={\frac{\Delta _{o}-\mu }{w\left[ {\frac{\Delta _{o}-\mu _{o}}{
\varepsilon }}\right] }}\sim {\frac{\Delta _{o}-\mu _{o}}{\ln \left[ {\frac{
\Delta _{o}-\mu _{o}}{\varepsilon }}\right] }}\,.
\end{equation}
It is then clear that the ratio $T_{c}/\Delta (T=0)$ diverges for $
\varepsilon $ going to zero. One should
note that the behaviour found for both $\Delta (T=0)$ and $T_{c}$ as a
function of $\tau $ is exactly the same of that found for these two
quantities as a function of $\varepsilon $. The origin is the same since the
gap at $T_{c}$ in this case becomes $\Delta _{o}-\mu $.

\begin{figure}[tbp]
\centerline{\psfig{file=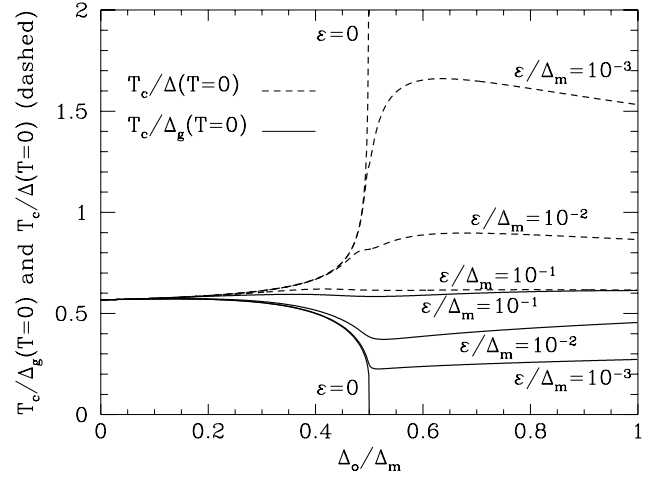,width=9cm,angle=-90}}
\caption{$T_{c}/\Delta _{g}(T=0)$ (continuous line) and $T_{c}/\Delta (T=0)$
(dashed line) for different values of $\varepsilon /\Delta _{m}$ 
(indicated on the plot). The lines corresponding to
$\varepsilon /\Delta _{m}=10^{-1}$ are nearly constant at the BCS value
of $\approx 0.57$. }
\label{fig4}
\end{figure}

We report in Fig.~\ref{fig4} the ratio between the critical 
temperature $T_{c}$ and the other two relevant
energy scales, the zero temperature order parameter $\Delta (T=0)$ and
excitation gap $\Delta _{g}(T=0)$ 
for different values of $\varepsilon /\Delta _{m}$ going from 0 to $
10^{-1}$.
For small $\Delta _{o}/\Delta _{m}$ the familiar BCS results of $e^{\gamma
}/\pi \approx 0.57$ is recovered, while for $\Delta _{o}>\Delta _{m}/2$ a
totally different behaviour is found. The critical temperature appears to be
in between the other two quantities (for $\varepsilon \ll \Delta _{m}$)
\begin{equation}
\Delta (T=0)<T_{c}<\Delta _{g}(T=0)\,.
\end{equation}
This can be easily checked analytically, since for $\varepsilon $ going to
zero $\Delta _{g}$ stays finite [$\Delta _{g}\rightarrow \Delta _{0}-\sqrt{
\Delta _{o}^{2}-\Delta _{m}^{2}/4}$], while $T_{c}$ vanishes logarithmically
and $\Delta (T=0)$ vanishes like $\sqrt{\varepsilon }$. Thus at mean field
level we have found a rather peculiar form of superconductivity, whose
behaviour is totally different from the BCS one. 
We note that the BCS behavior is recovered also for large values of 
$\Delta_o$ when the chemical potential lies in the conduction band. 
In Fig.~\ref{fig4} 
all the lines will join the BCS value 0.57 for {\em fixed} $\varepsilon$ 
and large enough $\Delta_o$. 
\begin{figure}
\centerline{
\psfig{file=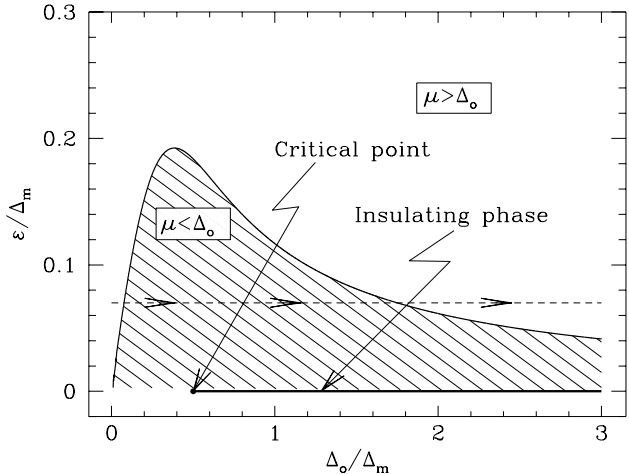,width=9cm,angle=-90}
}
\caption{Phase diagram for the chemical potential. The shaded part is
the region with $\mu<\Delta_o$. The thick line represents the 
insulating region of the phase diagram, for 
any other value of $\varepsilon$ and $\Delta_o$ 
the system is superconducting. The dashed line represents the path 
in the phase diagram described in the text.}
\label{phase}
\end{figure}
If $\varepsilon/\Delta_m$ is small enough, the parameter $\Delta_o$ 
drives the system through four different regimes. For 
very small $\Delta_o$ the system is in a two-band BCS regime. 
Increasing $\Delta_o$ the chemical potential enters the gap 
and the smoothed transition from a normal superconductor to a 
``semi-conductor'' superconductor takes place. 
When $\Delta_o \gg \Delta_o^*$ and the chemical potential is still 
in the gap, the carriers in the upper band act as a strong coupling 
superconductor, while the lower band is nearly 
decoupled. For even larger $\Delta_o$($>{\Delta_o^*}^2/2\varepsilon$) 
the chemical potential lies again in the valence band and a BCS 
behaviour is recovered for the upper band.  
The schematic phase diagram is plotted in Fig.~\ref{phase} 
where the region of positive and negative $\mu-\Delta_o$ are identified
on the plane $\Delta_o$-$\varepsilon$.
The path described above correspond to the dashed line.

All that discussion is based on a mean field, BCS like formulation.\ We
should now ask whether such an approximation is warranted. In the next
section we inquire about the effect of phase fluctuations in two dimensions,
which are known to provide an alternate mechanism for the critical
temperature $T_{c}$, based on the creation of vortices instead of pair
breaking.

\section{Phase stiffness and Kosterlitz$-$Thouless fluctuations }

\label{BKT}

We limit our discussion to a two dimensional system for which they 
are known to be crucial, leading to the so called Kosterlitz-Thouless 
transition \cite{BKT}. 
Although there is no strict long range order in 2d, the low temperature
decay of the order parameter is algebraic up to a sharp critical temperature
$T_{c}^{KT }$ at which the decay becomes exponential: such a transition
from a ``quasi superfluid'' phase to a normal phase is due to the unbinding
of vortex pairs. In the most naive approximation one may assume that the
modulus of the order parameter is constant, the only relevant variable being
its phase $\phi (\mathbf{x}).$ The corresponding fluctuations are controlled
by the phase stiffness $J$ defined by the following term in the Hamiltonian
\begin{equation}
\Delta H={\frac{J}{2}}\int d^{2}x|\nabla \phi (\mathbf{x})|^{2}
\end{equation}
$J$ measures the resistance of the system to phase gradients. From $J$ one
infers in 2d the well known exact critical temperature \cite{KT}
\begin{equation}
T_c^{KT}=\frac{\pi }{2}J \,.
\label{universalTc}
\end{equation}
The stiffness that enters (\ref{universalTc}) is actually temperature 
dependent, $J(T)$. That dependence occurs both at the mean field level,
through pair breaking thermal fluctuations, and via thermal fluctuations
of the phase itself, which can be described within a renormalization
scheme. We disregard the latter and we discuss briefly the mean field 
$J^{MF}(T)$ which provides a neat picture of the competition between 
BCS and phase fluctuations \cite{crossover}. 
$J^{MF}$ vanishes at the mean field transition
temperature $T^{MF}_c$, hence two possibilities:
({\em i\/}) $J^{MF}(0)\gg T_c$: phase fluctuations are essentially
negligible except very close to $T_c$: then $T_c^{KT}(T)$ becomes 
comparable to $T$ and BCS turns into a Kosterlitz-Thouless regime;
the ultimate transition is definitely in the Kosterlitz-Thouless 
universality class, but the relevant range is narrow and the real $T_c$ 
is very close to $T_c^{MF}$. Basically BCS is correct.
({\em ii\/}) $J^{MF}(0)\ll T_c$: then the Kosterlitz-Thouless transition
occurs well before pair breaking fluctuations become relevant: but for 
renormalization corrections $T_c$ is $T_c^{KT}(0)$.

What we need is an estimate of the stiffness $J(0)$. That is easily achieved
considering a superfluid flow in which pairs are condensed with center of
mass momentum ${\bf Q}/2$ instead of ${\bf 0}$. The corresponding 
phase of the order
parameter is $\phi ={\bf Q} \cdot {\bf r}$ and its gradient is 
${\bf Q}$. In order to calculate the
energy of that flow we need a specific model of the band structure in
momentum space: the knowledge of the density of states is not enough.\ Let
us first consider the standard model of a 2d free electron gas, with no gap
at all.\ The Fermi surface is a sphere with radius $k_{F},$ the density of
electrons with either spin is $N=k_{F}^{2}/2\pi .$ For a parabolic
dispersion law, with mass $m,$ the center of mass and relative motion of a
pair are completely decoupled: the cost in kinetic energy of the superfluid
flow is $N$ $
\rlap{\protect\rule[1.1ex]{.325em}{.1ex}}h%
^{2}Q^{2}/8m.$ The phase stiffness is consequently
\[
J=\frac{N
\rlap{\protect\rule[1.1ex]{.325em}{.1ex}}h%
^{2}}{4m}=\frac{\varepsilon _{F}}{4\pi }
\]
The corresponding transition temperature is $T_c^{KT}=\varepsilon
_{F}/8 $, comparable to the Fermi energy, but noticeably smaller due to the
accumulation of numerical factors. That must be compared to the BCS $
T_{c}\approx .57$ $\Delta $; as mentioned in the introduction the crossover
occurs when $\Delta $ is about a fifth of $\varepsilon _{F},$ making the
idea of a preformed bound state meaningless. Usual low temperature
superconductors are way below that crossover and phase fluctuations are
accordingly negligible.

We now turn to our gapped semiconductor: we need to specify the detailed
form of the two bands. The simplest guess assumes valence and conduction
bands that have a well defined extremum in the Brillouin zone, whether at
the same place (direct gap), or at different places (indirect gap). Near the
band edge the dispersion is parabolic, characterized by two (different)
masses (the values of these masses will not enter in the final expression of
the critical temperature). In a 2d system the density of states at band edge
is finite, and the assumption of a constant $\rho $ is not crucial: up to
that point, our model raises no difficulty. It does however when we try to
calculate the phase stiffness $J$. 
Before attempting explicit calculations,
let us explain the point qualitatively. In order to set up a superfluid flow
in a free electron gas, we shifted the Fermi surface in momentum space,
thereby obtaining a superfluid density $n_{s}= N$ and a phase stiffness 
$\sim \varepsilon _{F}$. If we do the same in a filled band we get nothing! A
filled band is a single state with no degree of freedom, and shifting the
distribution $n_{k}=1$ produces the same state. Hence the obvious
conclusion: \textit{a filled band has no stiffness.} $J$ can only result
either from doping or from the order parameter $\Delta $ that creates free
carriers even if there were none to start from. As a result the phase
fluctuation $T_c^{KT}$ is severely reduced: it becomes of order $
\Delta$, comparable to the BCS pair breaking $T_{c}$ in every case. The
difference with free carriers is dramatic, due to the narrow range of free
carriers in momentum space.

One more general comment is in order: is it possible to go smoothly from the
semiconductor to the metal? Put another way, does a reduction of the gap $
\Delta _{o}$ to $0$ drive us back to a regular Fermi liquid metal? As long
as we were dealing with the density of states $\rho$ only, the answer was
``yes'' in 2d: where free carriers sat in the Brillouin zone was of no
concern. When calculating the phase stiffness $J$ the answer is ``no''. The
standard situation is an indirect gap which closes when the minimum of the
conduction band is lower than the top of the valence band. Thereafter free
carriers exist even in the normal state, but they remain localized near the
corresponding band extrema. The insulator with no Fermi surface will not
turn suddenly into a metal with a large Fermi surface (a sphere for free
carriers). Instead a small Fermi surface will appear near the band extrema,
responsible for the very small superfluid density (and phase stiffness).
Such a behaviour should be contrasted with the 1d case, in which a smooth
transition is possible. The standard example is the appearance of a
commensurate spin density wave at zone edge: the resulting antiferromagnetic
order produces Bragg scattering and opens a gap $\Delta _{o}$.
Near the gap
the band dispersion is hyperbolic, returning to the normal metal behaviour
when the distance $( \varepsilon -\varepsilon _{0})$ 
to the band bottom $\varepsilon_o$ exceeds $\Delta _{o}:$
in such a case the transition metal $\rightarrow$ semiconductor is smooth.
But then the physics is completely different, as usual in 1d geometries. We
see no realistic 2d system that could provide a smooth transition and a
large superfluid density (that would mean a gap spread over the whole Fermi
surface and translating together with it).

We now turn to the calculation of $J$, putting all the pairs in a state of
total momentum $\mathbf{Q}$ and looking at the variation of the total
energy. The BCS calculation for the energy is unmodified but for the
following transformation
\begin{equation}
\xi _{k}\rightarrow \Xi _{k}(Q)=(\xi _{\mathbf{k}+\mathbf{Q}/2}+\xi _{-%
\mathbf{k}+\mathbf{Q}/2})/2\,.
\end{equation}
where $\xi _{\mathbf{k}}$ is the dispersion relation of the band. This
transformation changes only the kinetic energy: in our case we have for the
two bands:
\begin{equation}
K(Q)=2\sum_{\mathbf{k}\in v}v_{k}^{2}[\Xi _{\bf k}^{v}({\bf Q})-
\mu ]+2\sum_{\mathbf{k}
\in c}v_{k}^{2}[\Xi _{\bf k}^{c}({\bf Q})-\mu ]\,.
\end{equation}
Since the shift of $\mathbf{Q}$ of a filled band does not change the kinetic
energy of the band we can write:
\begin{eqnarray}
	K(Q) 
	&=&
	2 \sum_{\mathbf{k}\in v}(\xi _{k}^{v}-\mu )
	-2\sum_{\mathbf{k}\in v} u_{k}^{2} [\Xi _{\bf k}^{v}({\bf Q})-\mu ]
	\nonumber \\
	&& 
	+2\sum_{\mathbf{k}\in c} v_{k}^{2}
	[\Xi_{\bf k}^{c}({\bf Q})-\mu ]\,.
\end{eqnarray}
We use at this point the hypothesis that in the region where $u_{k}$ and $
v_{k}$ are significantly different from zero the two bands are parabolic. We
can thus expand $\Xi $ for small $Q$:
\begin{equation}
\Xi _{\bf k}^{v}({\bf Q})\simeq \xi _{k}^{v}-{\frac{Q^{2}}{8m_{v}}}\quad 
\mathrm{and}
\quad \Xi _{\bf k}^{c}({\bf Q})\simeq \xi _{k}^{c}+{\frac{Q^{2}}{8m_{c}}}
\end{equation}
where $m_{c/v}$ are the effective masses of the two bands. The expression
for $K$ becomes:
\begin{equation}
K(Q)=K(0)+{\frac{Q^{2}}{4m_{v}}}\sum_{\mathbf{k}\in v}u_{k}^{2}+{\frac{Q^{2}%
}{4m_{c}}}\sum_{\mathbf{k}\in c}v_{k}^{2}
\end{equation}
By introducing the 2-dimensional density of states per spin: $\rho ={\
m/2\pi }$ we obtain:
\begin{equation}
K(Q)-K(0)={\frac{Q^{2}}{8\pi }}\left[ \int_{-\infty }^{-\Delta _{o}}u(\xi
)^{2}d\xi +\int_{\Delta _{o}}^{+\infty }v(\xi )^{2}d\xi \right] \,,
\end{equation}
and the masses simplify from the expression. We thus obtain for $J$:
\begin{equation}
J=\int_{\Delta _{o}}^{+\infty }\!\! {d\xi \over 8\pi}
	\left[ 2-{\frac{\xi +\mu }{%
	\sqrt{(\xi +\mu )^{2}+\Delta ^{2}}}}-{\frac{\xi -\mu }{\sqrt{(\xi -\mu
)^{2}+\Delta ^{2}}}}\right] \,\,,  \label{stiffness}
\end{equation}
and since the integration is elementary we can write a closed expression for
$T_c^{KT}$
\begin{equation}
T_c^{KT}=
{\sqrt{(\Delta _{o}-\mu )^{2}+\Delta ^{2}}+%
\sqrt{(\Delta _{o}+\mu )^{2}+\Delta ^{2}}-2\Delta _{o}
\over 16}
\,.
\label{stiff}
\end{equation}
We comment this result in the two cases of $\varepsilon =0$ and of finite
doping.

In the undoped case the expression simplifies even more. 
In the superconducting region, $\Delta_o<\Delta_o^*$, we find 
\begin{equation}
T_c^{KT} = (\Delta_m-2\Delta_o)/8
\label{eq38}
\end{equation}
$T_c^{KT}$ is thus comparable to the order parameter $\Delta$, 
in contrast to a BCS superconductor, a spectacular consequence of the 
severe reduction in phase stiffness. The behavior near the transition
point is easily understood. There the gap is large as compared  to 
$\Delta$ and the density of free carriers $v_k^2$ lies in the tail 
of the BCS distribution: it is of the order $\rho \Delta^2/\Delta_m$.
The stiffness is accordingly of the same order, leading to (\ref{eq38}).
Our conclusion is that the transition to the normal state is always 
due to unbinding of vortices, in sharp contrast to the weak coupling BCS 
limit.

We consider now the case of large $\Delta_o$. 
We can use the approximation used in (\ref{system2}) and keeping the lowest
order in $\Delta/\Delta_o$ in Eq.~(\ref{stiff})
we find:
\begin{equation}
T_c^{KT} = 
\frac{\varepsilon }{8} 
\left[1+{{\Delta_o^*}^2\over 2 \Delta_o^2}\right]
\,.
\end{equation}
As expected $T_c^{KT}$ is proportional to the stiffness of a single band with 
Fermi energy $=\varepsilon$, this result is true for any sign of 
$\mu-\Delta_o$ provided $\mu\sim\Delta_o\gg\Delta$.
The factor of proportionality is greater than
one indicating that the lower band is contributing to the stiffness.
This result implies that even in the doped system 
(for $\varepsilon\rightarrow 0$) the lowest critical temperature is $T_c^{KT}$ 
due to binding-unbinding transitions of the vortices. 
The anomalous large value of
the critical temperature found at mean field is thus never reached in
reality. This result is not changed by a more realistic model for the
dispersion relation as shown in Appendix \ref{AppA}.

\section*{Conclusions}

Superconductivity is usually concerned with a regular metal, in which free
carriers exist at the Fermi level. We show here that superconductivity can
also occur in semiconductors as soon as carriers in either band feel an
attraction. The semiconductor is characterized by a gap $2\Delta _{o},$ the
origin of which is not specified. The attraction is characterized by the
superconducting gap $\Delta _{m}$ that would exist in a regular metal. The
relevant parameters are the ratio $\Delta _{o}/\Delta _{m}$ and the doping
of the semiconductor.

Superconductivity may happen even in an undoped semiconductor in which no
free carriers exist in the normal state at $T=0$, as soon as the gain in
superconducting energy $\sim \Delta _{m}$ exceeds the cost in producing
carriers $\sim \Delta _{o}$. Such a situation is completely equivalent to
excitonic insulators, in which electrons repel instead of attracting: then
electrons and holes attract and they form bound excitons.\ If the binding
energy is larger than the gap these excitons appear spontaneously. In our
model we find a sharp transition at a specific value $\Delta _{o}^{*}$ $\sim
\Delta _{m}$ (the precise value depends on details of the density of
states). The ground state is superconducting if $\Delta _{o}<\Delta _{o}^{*},
$ insulating if $\Delta _{o}>\Delta _{o}^{*}$.  
Near the transition it departs
from the familiar BCS regime: the order parameter $\Delta $, the real
quasiparticle gap  $\Delta _{g}$ and the critical temperature $T_{c}$ are
such that
\[
\Delta \ll T_{c}\ll \Delta _{g}
\]
That may be viewed as a ``pseudogap'' behaviour, monitored by the
competition between semiconducting and superconducting features.

In the undoped case only the vicinity of the transition point $\Delta
_{o}^{*}$ is unusual. This is no longer true when the semiconductor is doped:
then the sharp transition becomes a crossover and superconductivity extends
all the way. Nothing very spectacular happens when $\Delta _{o}<\Delta
_{o}^{*}:$ the effect of a small doping is minor. In contrast the
``semi-insulating region'' $\Delta _{o}>\Delta _{o}^{*}$ is extremely
unusual: that is the main result of our paper. Past the crossover the
chemical potential first remains inside the band gap. We then find a
superconducting state very reminiscent of the Bose Einstein condensation of
preformed pairs. The order parameter $\Delta $ is proportional to the square
root of doping, and the characteristic energies display the same pseudogap
ordering as in \cite{NSR}. 
That behaviour prevails until the gap reaches $\Delta
_{o}^{*2}/\varepsilon ,$ where $\varepsilon $ is a typical Fermi energy in
the conduction band.\ Then the chemical potential enters the conduction band
and, after a transient region, the system returns to a single band BCS
superconductor that involves only extrinsic carriers. The range of anomalous
behaviour is very broad, and thus generic.

The main body of the paper is concerned with the simplest conceivable model,
with a constant density of states $\rho $ (a sensible choice for a 2d
system), and a mean field treatment ``\`{a} la BCS'', both at zero and at
finite temperatures. The only new feature is the presence of the
semiconductor gap $\Delta _{o}.$ The problem is then studied both
analytically and numerically. Both simplifications can be questioned: we now
address the issues in succession.

A constant density of states is not realistic for two reasons: ({\em i\/})
 it ships
the gap states to infinity, while standard mechanisms for a small gap leave
the states close to where they originated from, in an energy range $\sim
\Delta _{o}$ ({\em ii\/}) 
in 3d the density of states $\rho \left( \xi \right) $
starts as $\sqrt{\xi }$ near band edge. Improvements are explored in
Appendix \ref{AppA}, 
where other densities of states are tried, still within a mean
field picture. Numerical results show that the qualitative evolution remains
the same.\ In the undoped case the sharp superconductor-insulator transition
is just shifted, the square root behaviour of $\Delta $ being unaffected. In
the doped case the calculation can be carried out explicitly in the
anomalous region, if
\[
\Delta \ll \Delta _{o}-\mu \ll \Delta _{o}
\]
Free carriers at $T=0$ are then mostly in the conduction band and their
density $v_{k}^{2}$ is small, of order $\Delta ^{2}.$ Our main result, $%
\Delta $ proportional to the square root of doping, is maintained (with
modified coefficients). We conclude that the physics described in this paper
is not too sensitive to details of $\rho \left( \xi \right)$. We did not
consider the case of a Fermi surface close to a Van Hove singularity, as
found in a nested 2d system.\ That could definitely be done.

Departures from mean field behaviour is a more delicate problem. Thermal
fluctuations of the phase of $\Delta $ are controlled by the phase stiffness
$J.$ In ordinary metals $J$ is comparable to the Fermi energy: phase
fluctuations are small and a mean field picture is correct except very close
to $T_{c}^{MF}$ (where fluctuations turn to a Kosterlitz-Thouless transition
in 2d). In our semiconductor we show that $J$ is only due to free carriers
(it would be $\equiv 0$ for a filled band). $J$ is comparable to the Fermi
energy and thus small: it follows that the phase fluctuation mechanism for 
$T_{c}$ always prevails 
(at least in 2d and for $\Delta_o<\Delta_m^2/\varepsilon$): 
that is the second unexpected result of our analysis. 
Simple considerations are enough in order to provide
orders of magnitude, but a real quantitative theory is still lacking.

Granted these difficulties, we only tackled the problem, our goal being to
emphasize simple physical ideas. Many important questions remain open.\ For
instance one should worry about the origin of the semiconducting gap $\Delta
_{o}.$ If it arises from another instability (lattice distortion, magnetism,
etc...), does superconductivity react on that primary instability? If it
does we face a coupled problem, in which $\Delta $ depends on $\Delta _{o}$
and in reverse $\Delta _{o}$ depends on $\Delta .$ A self consistent
calculation is required, and it may lead to unexpected 
fixed points \cite{imada}. 
Another issue is the relevance of interband pairing, as envisaged
 in \cite{kohmoto}. Even if it is not the primary effect, it may affect
the quantitative results. (It seems that such am interplay of intra and 
interband pairing is crucial in understanding the smooth transition
of the stiffness when the gap $\Delta_o$ closes to zero in a 1D band.)
These remarks open exciting possibilities, 
which we did not approach. As it stands our
paper is only an exploration.

\section*{Acknowledgments}

This work evolved from long discussions with Guy Deutscher: we are 
grateful to him for his helpful advice. We are also grateful to 
Prof.~J.~Friedel for pointing out Ref.~\cite{kohmoto} of which we 
were not aware.


\appendix

\section{Non constant density of states}

\label{AppA}

In this appendix we consider a more realistic form for 
the density of states, in which the states removed from the
energy region $[-\Delta_o,\Delta_o]$ remain in the vicinity of the edge. 
We explore two simple cases: a discontinuous density of states 
(typical of 2d system) and a density of states that 
vanishes at $\Delta_o$ with a square root of $\xi-\Delta_o$ 
(typical of a 3d case). 
These two possibilities are shown in Fig.~\ref{dos}. 
We solve the equations numerically
showing that the qualitative behaviour found for constant density 
of states is not changed. 

Eqs.~(\ref{eqA}) and (\ref{eqB}) can be easily generalized 
when the density of states is not constant:
\begin{equation}
\left\{
\begin{array}{l}
\displaystyle \int_{P^{\prime}} \rho(\xi) \left[ v^2(\xi) + f_F[E(\xi)] {%
\frac{\xi-\mu }{E(\xi)}} \right] d\xi = \int_{-\omega_m}^{\varepsilon}
\rho_m(\xi)\, d\xi \\
\displaystyle \int_{P^{\prime}} \rho(\xi) {\frac{\tanh \beta E(\xi)/2 }{
E(\xi)}} \,d\xi = \int_{-\omega_m}^{+\omega_m} {\frac{\rho_m(\xi) }{\sqrt{
\Delta_m^2+\xi^2}}} \, d\xi
\end{array}
\right. \,.  \label{sysrho}
\end{equation}
In this equations $\rho_m(\xi)$ and $\rho(\xi)$ are the densities of states
of the metallic and insulating phase respectively. The integration path is now
 $P^{\prime}=[-\omega_m,-\Delta_o]$ and $[\Delta_o,+\omega_m]$, since 
we conserve the number of states by deforming $\rho$ and not by
shifting the whole band. 
The function $\rho(\xi)(\geq 0)$ is thus constrained by the 
conservation of the number of states when passing from the metallic 
to the insulating phase:
\begin{equation}
\int_{P^{\prime}} \rho(\xi) \,d\xi = \int_{-\omega_m}^{\omega_m} \rho_m(\xi)
\, d\xi \,.  \label{up}
\end{equation}
It is clear that in order to fulfill Eq.~(\ref{up}) 
$\rho(\xi)$ must depend on $\Delta_o$ too.

\begin{figure}
\psfig{file=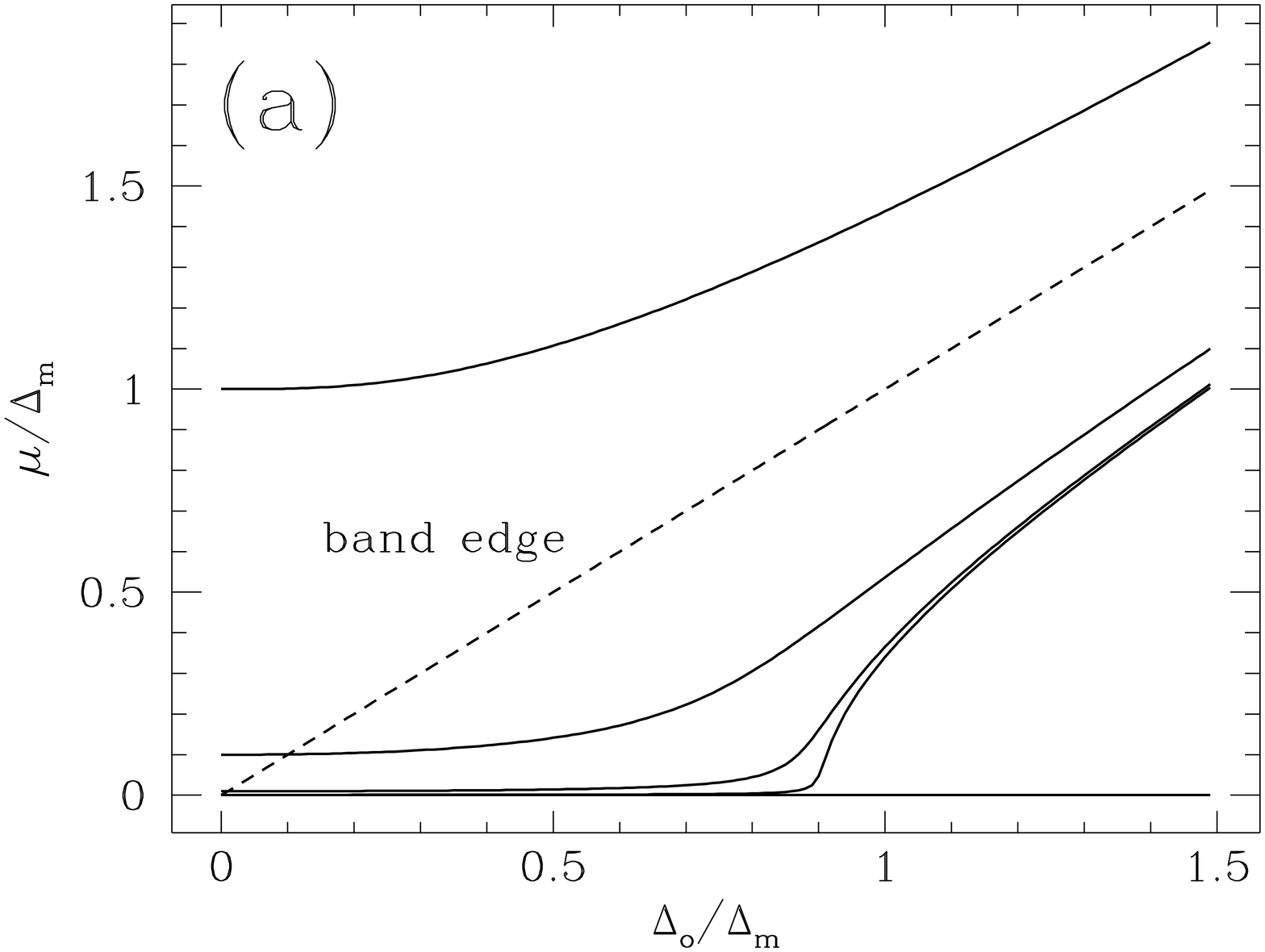,width=9cm,angle=-90}
\psfig{file=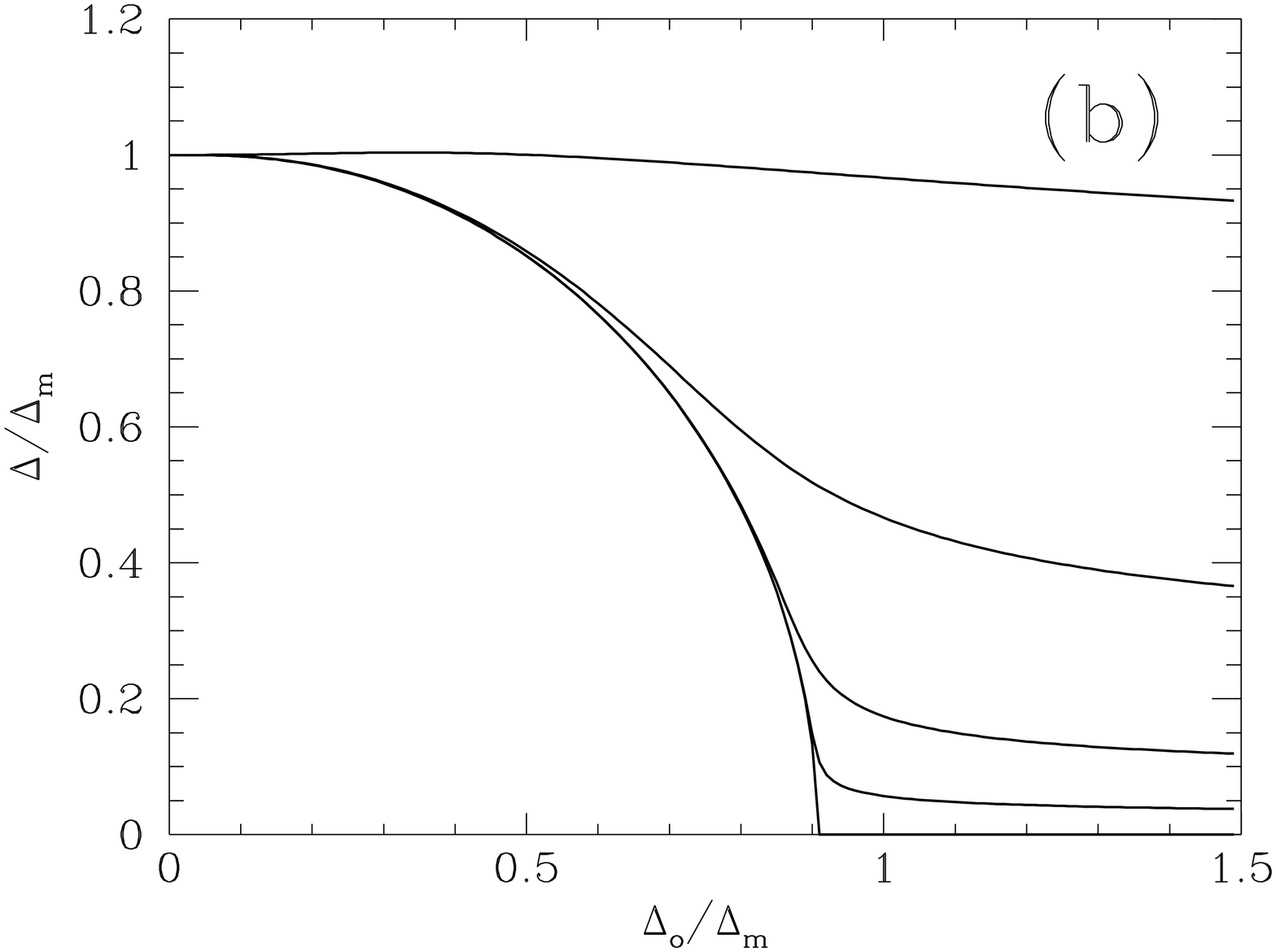,width=9cm,angle=-90}
\caption{Zero temperature numerical results for the density of states 
$\rho=\rho_1$ (discontinuous one): 
(a) Chemical potential, and bottom of the upper band (dashed).
(b) Order parameter $\Delta$. $\varepsilon/\Delta_m$ take the following
values: 0, $10^{-3}$, $10^{-2}$, $10^{-1}$, 1. 
}
\label{figA1}
\end{figure}

\begin{figure}
\psfig{file=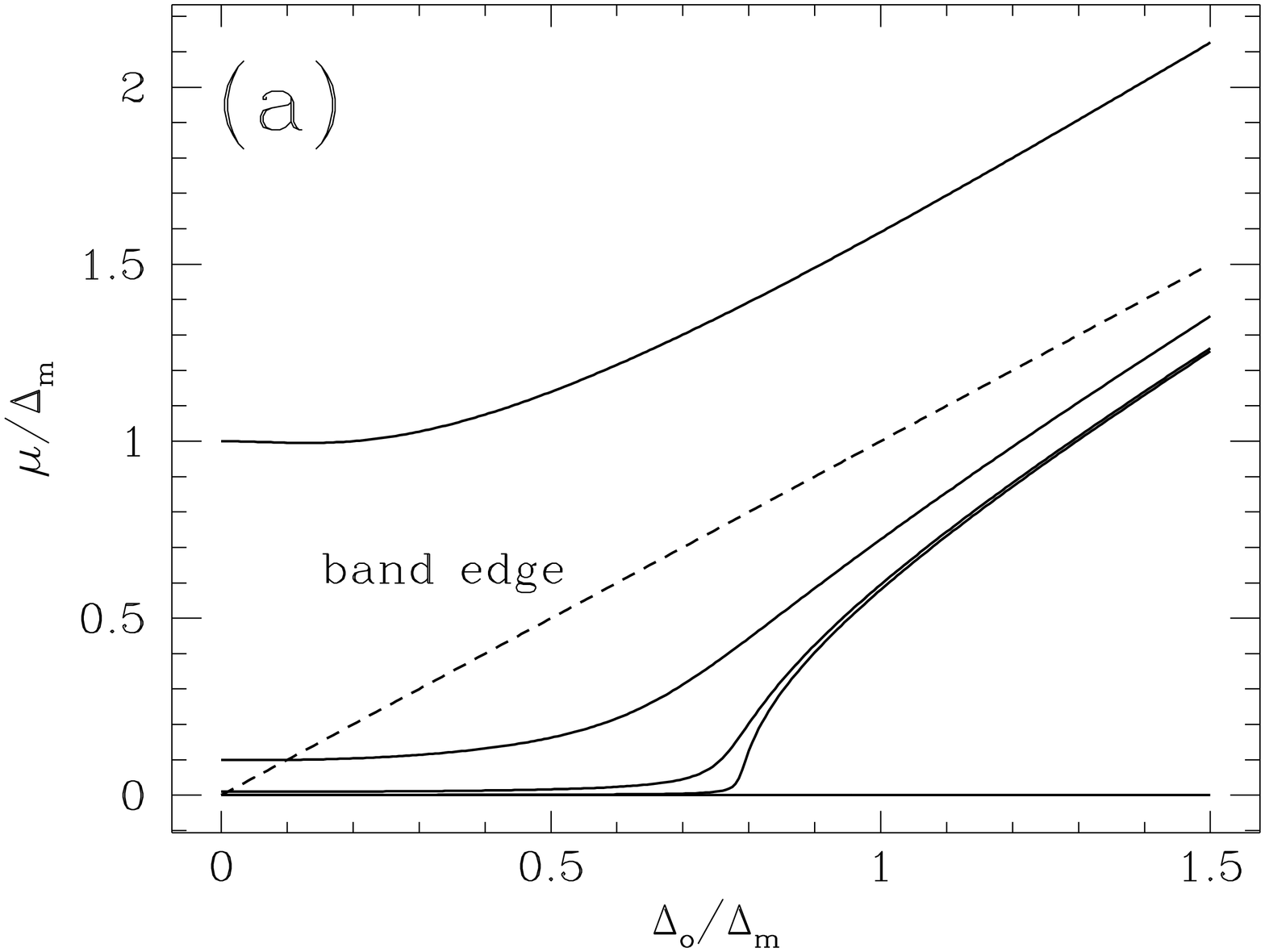,width=9cm,angle=-90}
\psfig{file=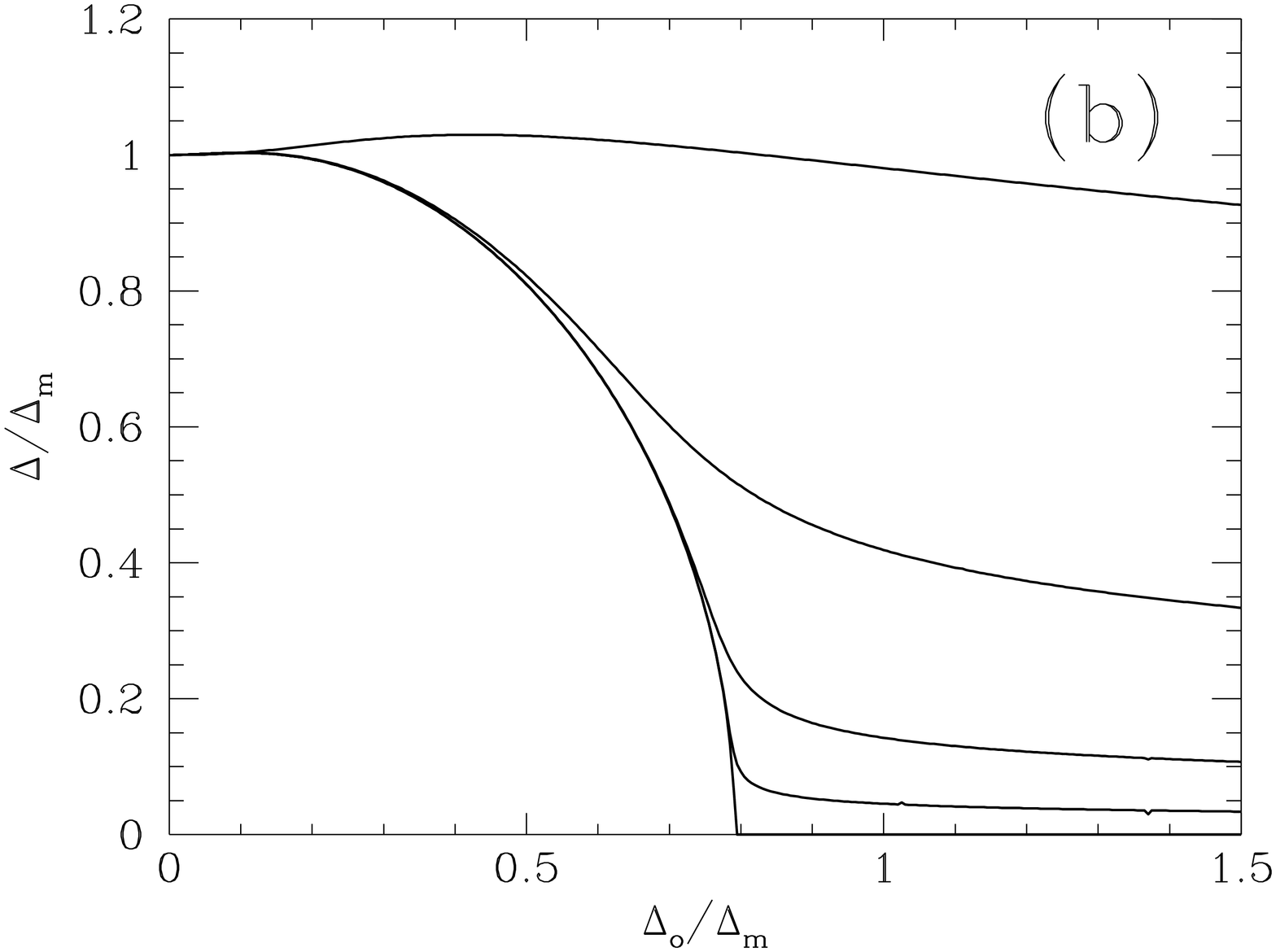,width=9cm,angle=-90}
\caption{Numerical results for the density of states 
$\rho=\rho_2$ (square root). Conventions are the same of Fig. \ref{figA1}. 
}
\label{figA2}
\end{figure}

Equations (\ref{sysrho}) are quite general. We will consider two 
specific simple cases. We assume that $\rho_m$ is constant, and that 
$\rho(\xi)$ is defined in the two cases by the following expressions:
\begin{equation}
\left\{
\begin{array}{rcl}
	\rho_1(\xi)/\rho_m
	\displaystyle 	
	&=& 1 + e^{-x}\\
	\displaystyle 
	{\rho_2(\xi)/\rho_m} 
	&=& \displaystyle 
	{2+\pi \over\sqrt{\pi}} \sqrt{x}\, e^{-x} + {x^2\over 1+x^2}
\end{array}
\label{simp}
\right.
\end{equation}
where $x=(|\xi|-\Delta_o)/\Delta_o$ and $\rho=0$ for 
$|\xi|<\Delta_o$ [cf. Fig.~\ref{dos}, the dotted line correspond to 
$\rho_1$ and the dashed line to $\rho_2$].
The numerical results for $\Delta$ and $\mu$ in the zero temperature 
case are reported in 
Fig.~\ref{figA1} for the discontinuous density of states ($\rho_1$),
and in Fig.~\ref{figA2} for the second one.
There is no qualitative difference with Fig. \ref{fig2}.
Different shapes of the density of states correspond to
different values of the ratio $\Delta^*/\Delta_m$,
always keeping it of order 1\footnote{
Although the energy range over which $\rho(\xi)$ varies is of 
order $\Delta_o$, we have tried a more general form in which 
that energy scale is different  from $\Delta_o$.
The maximal variation of $\Delta_o^*$ obtained is an
increase of a factor $e \approx 2.71$ when all the removed states 
are accumulated at the edge. 
}.

Most of the results can be understood analytically.
Let us first consider how 
the order parameter vanishes at the transition for no doping. 
In this case $\Delta$ is 
small and we can expand $E$ in the second of (\ref{sysrho}). 
After the equation for the critical $\Delta_o$ is subtracted
we obtain:
\begin{eqnarray}
	0 &=&  
	\int_{\Delta_o^*}^{+\infty} 
	{\rho_{\Delta^*_o}(\xi)-\rho_{\Delta_o}(\xi) \over \xi}
	d\xi
	\nonumber \\ 
	&& 
	-{1\over \Delta_o} \int_{\Delta_o}^{\Delta_o^*} \rho(\xi)d\xi
	+\Delta^2 
	\int_{\Delta_o^*}^{+\infty}{ \rho_{\Delta_o} \over 2 \xi^3}
	\,.
\end{eqnarray}
The last term for $\tau\rightarrow 0$ ($\Delta_o\rightarrow \Delta_o^*$) 
yields simply $\Delta^2(\tau)$ multiplied by a constant. 
The second term vanishes like $\tau^{1+\alpha}$ 
if $\rho(\xi) \sim (\xi-\Delta_o)^\alpha$. This
term is always subleading compared to the first term that 
vanishes linearly in $\tau$. We thus conclude that the dependence 
$\Delta \sim \tau^{1/2}$ is quite general. 
This is not surprising for a mean field theory.

For finite doping and $T=0$ it is interesting to study the dependence 
of $\Delta$ on $\varepsilon$. 
In particular we can readily obtain this
dependence analytically when $\Delta_o \gg \Delta_m$. 
We distinguish the two cases $\mu<\Delta_o$ and $\mu>\Delta_o$. When 
$\Delta_o \gg \Delta_o -\mu \gg \Delta$ the lower band is nearly
decoupled and the conservation of particles gives:
\begin{equation}
	\delta N = 2 \varepsilon \rho_m  
	\approx 
	\int_{\Delta_o}^{\infty} \rho(\xi) {\Delta^2 \over 2 (\xi-\mu)^2}
	\approx
	{\Delta^2 \over 2} 
	 \, { \rho(\Delta_o-\mu) \over \Delta_o-\mu}
\end{equation}
We thus recover $\Delta\sim \varepsilon^{1/2}$ for a general density of 
states, again a mean field exponent. 
Note that the coefficient involves the density of states on an 
energy scale $\sim (\Delta_o-\mu)$ 
(measured from the bottom of the conduction band).
This fact does not invalidate the conclusion since the chemical
potential is practically independent on $\varepsilon$, determined 
as usual by the gap equation. 

For $\Delta_o$ large enough $\mu$ sits again in the valence band,
and the limit $\Delta_o \gg \mu-\Delta \gg \Delta$ is reached.
In this case the BCS theory applies in the upper band and 
one recovers the familiar exponential behavior of $\Delta$ 
on the inverse of the density of states at the Fermi energy.

Concerning the critical temperature we have verified that using 
the density of states $\rho_1$ one recovers the same behaviour 
found for a constant density of states: near the transition point
and for small doping it vanishes like 
$1/w(1/\tau)$ or $1/w(1/\varepsilon)$.
In the BCS region ($\Delta_o\gg \Delta_m^2/\varepsilon$) 
the usual BCS $T_c$ is recovered.

We can readily calculate the phase stiffness if we assume that 
the dispersion relation depends only on ${\bf k}^2$.
In this case we can generalize the results for the variation 
of the kinetic energy: 
\begin{equation}
\Xi_\mathbf{k}(\mathbf{Q}) = \xi_k + \xi_k^{\prime}{\frac{Q^2 }{4}} +
\xi_k^{\prime\prime}{\frac{(\mathbf{k} \mathbf{Q} )^2 }{2}} +O(Q^4)
\,,
\end{equation}
where $\xi' \equiv d\xi(k^2)/d(k^2)$.
Using this expression for the kinetic energy and the invariance under
rotation we obtain for each band:
\begin{equation}
{\frac{d E(Q) }{d Q^2}} = 2 \sum_{\mathbf{k}\in \mathrm{band}} 
v_k^2\left[{\frac{
\xi_k^{\prime}}{4}}+{\frac{\xi_k^{\prime\prime}}{2}} {\frac{k^2}{2}}\right]
\,.
\end{equation}
The derivatives of the dispersion relation can be related to the density of
states: $\xi_k^{\prime}=1/4\pi \rho$ and $\xi_k^{\prime\prime}=
-\rho^{\prime}/16 \pi^2 \rho^3$. Putting everything  together 
we obtain for our
problem the same phase stiffness $J$ as in (\ref{stiffness}), but with
an additional factor $A(\xi)$ in the integrand of (\ref{stiffness})
[provided $\rho(\xi)=\rho(-\xi)$]:
\begin{equation}
A(\xi) = 1 - {\frac{{\rho}^{\prime}(\xi) }{{\rho(\xi)}^2}}
\int_{\Delta_o}^\xi \rho(x) \, dx \,.
\end{equation}
When the density of states is constant $A(\xi)$ is 1. 
One can show that the integral of $[A(\xi)-1]$ is equal to $\Delta_o$: 
it thus gives the contribution of the additional states to the phase 
stiffness. 
The various contributions to the phase stiffness
are just rearranged and we find that the shape of $\rho(\xi)$  
does not affect
qualitatively the dependence of the phase stiffness on the 
important parameters $\Delta_o$ and $\varepsilon$. 
Moreover the quantitative change is not enough to let $T_c^{KT}$ 
become larger than the $T_c$ given by particle excitations. 
Thus the phase fluctuations should always be 
responsible for the transition to the normal phase.
(Except for very large $\Delta_o$ at fixed $\varepsilon$ 
where the BCS limit is recovered.)

\section{Interband pairing}

\label{AppB}

In this paper we have considered  pairing of time reversed state
within the same band. If we indicate with $a_{k\sigma}$ and 
$b_{k\sigma}$ the 
destruction operators for the lower and upper band respectively, 
the possible pairings considered in the main body of the paper are:
$
	\langle a_{k\uparrow} a_{-k\downarrow}\rangle
$
and
$
	\langle b_{k\uparrow} b_{-k\downarrow}\rangle
$. The two bands are then coupled by the interband interaction,
 which is assumed identical to the intraband interaction. 

Another model of pairing in a two band semiconductor 
was proposed some years ago by Kohmoto and Takada \cite{kohmoto}
(see also \cite{mhk}). The main difference with our proposal 
is that in that case the paired states are not time-reversed one 
of the other, specifically electrons of different bands are paired: 
$
	\langle a_{k\uparrow} b_{-k\downarrow}\rangle
$.
(The spin was not included in the original formulation and 
it is introduced here for comparison with our work).
The mean field calculations predicts at $T=0$
a first order transitions from an insulator to a superconductor.
We discuss briefly in this Appendix this model \cite{kohmoto}
and we compare it to ours in one simple case.

The main difference arise from the fact that the kinetic 
energies of the paired states are different: one expects that the 
energy cost should accordingly be larger.
Although it is possible to have interactions that favor 
only one of the two possible states, 
when the strength of the interaction is the same in the 
two channels the intraband pairing should be favoured 
due to the lower (kinetic) energy cost. 
We demonstrate this fact in a simple case.

We consider a local interaction of the form
\[
-U \int \!dx\,\rho(x) \rho(x)\,,\] 
where 
$
	\rho(x) 
	= 
	\sum_\sigma \psi^\dag(x)_\sigma \psi^{\phantom{\dag}}(x)_\sigma 
$ and $\psi_\sigma(x)=a_\sigma(x)+b_\sigma(x)$.
The antiparallel spin component of this interaction can be reduced to 
two terms:
\begin{eqnarray}
	V_1 &=& -2U \int\!\! dx\, 
	\left(
	a^\dag_\uparrow a^\dag_\downarrow
	+
	 b^\dag_\uparrow b^\dag_\downarrow 
     	\right)
	\left( 
	a^{\phantom{\dag}}_\downarrow a^{\phantom{\dag}}_\uparrow
	+
	b^{\phantom{\dag}}_\downarrow b^{\phantom{\dag}}_\uparrow 
	\right)\,, \\
	V_2 &=& -2U \int \!\!dx\,
	\left(
	a^\dag_\uparrow b^\dag_\downarrow
	-
	a^\dag_\downarrow  b^\dag_\uparrow 
     	\right)
	\left( 
	b^{\phantom{\dag}}_\downarrow a^{\phantom{\dag}}_\uparrow
	-
	b^{\phantom{\dag}}_\uparrow a^{\phantom{\dag}}_\downarrow 
	\right)\,.
\end{eqnarray}	
$V_1$ would induce the pairing suggested in this 
paper, while $V_2$ is needed in order to obtain the state 
proposed by Kohmoto and Takada.
We consider the simplest case of constant density of states $\rho$,   
symmetric bands, and no doping ($\mu=0$). 
We also keep the  energy cut-off 
$\Delta_o+\omega_c=\omega_m$ finite.
We can then evaluate the Hamiltonian on the respective 
trial wave functions. For the intraband pairing proposed in this paper 
the ground state wavefunction is:
\begin{equation}
	|\psi_1\rangle = \prod_{\bf k} 
	\left(u_k+v_k \,b^\dag_{k\uparrow} b^\dag_{-k\downarrow} \right)
	\left(v_k+u_k \,a^\dag_{k\uparrow} a^\dag_{-k\downarrow} \right)
	|vac\rangle
\end{equation}
while for the state proposed in Ref.~\cite{kohmoto}
the wavefunction takes the form:
\begin{equation}
	|\psi_2\rangle = \prod_{\bf k} 
	\left({1+a^\dag_{k\uparrow} b^\dag_{-k\downarrow} 
	\over \sqrt{2} }
	\right)
	\left(
	{ 1- a^\dag_{k\downarrow} b^\dag_{-k\uparrow} 
		\over 
	\sqrt{2}}
	\right)
	|vac\rangle
	\,.
\end{equation}
Here the singlet state has been chosen to take maximal advantage of 
the interaction. 
The energy difference with respect to the normal state ($\delta E$) 
is in the two cases: 
\begin{eqnarray}	
	\delta E_1 &=& \rho\left[2\Delta_o \omega_c+\omega_c^2-
	(\Delta_o+\omega_c)
	\sqrt{(\Delta_o+\omega_c)^2+\Delta^2}
	\right.\nonumber \\
	&& \left.
	+ \Delta_o\sqrt{\Delta_o^2+\Delta^2}
	\right]
	\,,
	\label{en1}
	\\
	\delta E_2 
	&=&
	\rho\omega_c \left[ 
	2\Delta_o+\omega_c-2 U \omega_c \rho
	\right]
	\,,
\end{eqnarray}	
where the value of $\Delta$ in (\ref{en1}) is given by the usual gap equation:
$
	1 = 2 \rho U \int_{\Delta_o}^{\Delta_o+\omega_c}
	{d\xi/\sqrt{\xi^2+\Delta^2}}
$.
The minimum value of the interaction ($U^{min}$) 
for which $\delta E$ changes sign determines the 
transition from the insulating to the superconductig phase.
For the two states we have respectively
$\rho U^{min}_1= 1/2\ln(1+\omega_c/\Delta_o)$ and 
$\rho U^{min}_2= 1/2 +\Delta_o/\omega_c)$.
Thus $U^{min}_1<U^{min}_2$ always.  
Moreover  explicit evaluation of the energy shows that the 
energy gain of the intraband pairing state is always larger than 
the interband one.
\begin{figure}
\centerline{
\psfig{file=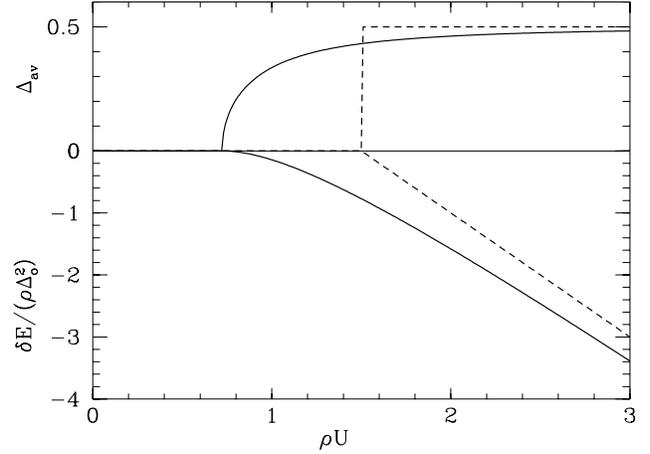,width=9cm,angle=-90}
}
\caption{Comparison of the ground state energies of the state presented 
in this paper (full lines) and of the one proposed 
in Ref~\cite{kohmoto} (dashed line). The upper part of the plot
shows the order parameter for the two models. 
}
\label{figB1}
\end{figure} 
In Fig.~\ref{figB1} we plotted  $\delta E$ in the 
two cases when $\delta \omega_c/\Delta_o=1$. 
The transition is second order in case (1), 
first order in case (2).
We report also the average of the 
order parameter over the states that participate to the pairing:
\begin{equation}
	\Delta_{av} = {\sum_k u_k v_k \over \sum_k 1} \,,
\end{equation} 
which clearly displays the nature of the transition.

\end{document}